\documentclass{jfm}

\usepackage[colorlinks,citecolor=blue,urlcolor=red,bookmarks=false,hypertexnames=true]{hyperref} 
\hypersetup{ 
     colorlinks=true, 
     linkcolor=blue, 
     filecolor=blue, 
     citecolor = blue,       
     urlcolor=blue, 
     } 
\usepackage{graphicx}
\usepackage{epstopdf, epsfig}
\usepackage{natbib}
\usepackage{amsmath}
\usepackage{dcolumn}
\usepackage{bm}
\usepackage{multirow}
\usepackage{multicol}
\usepackage{hyperref}

\title{A transformer-based synthetic-inflow generator for spatially-developing turbulent boundary layers}

\author{Mustafa Z. Yousif$^1$,  Meng Zhang$^1$, Linqi Yu$^1$, Ricardo Vinuesa$^2$ and  HeeChang Lim$^{1, \corresp{\email{hclim@pusan.ac.kr}}}$}

\affiliation{$^1$School of Mechanical Engineering, Pusan National University, 2, Busandaehak-ro 63beon-gil, Geumjeong-gu, Busan, 46241, Rep. of KOREA\\[\affilskip]
$^2$FLOW, Engineering Mechanics, KTH Royal Institute of Technology, Stockholm, Sweden}
\pubyear{2022}
\volume{650}
\pagerange{119--126}
\date{Accepted 15 December 2022. - accepted by editorial office}
\begin{document}

\maketitle

\begin{abstract}
This study proposes a newly-developed deep-learning-based method to generate turbulent inflow conditions for spatially-developing turbulent boundary layer (TBL) simulations. A combination of a transformer and a multiscale-enhanced super-resolution generative adversarial network is utilised to predict velocity fields of a spatially-developing TBL at various planes normal to the streamwise direction. Datasets of direct numerical simulation (DNS) of flat plate flow spanning a momentum thickness-based Reynolds number, $Re_\theta$ = 661.5 – 1502.0, are used to train and test the model. The model shows a remarkable ability to predict the instantaneous velocity fields with detailed fluctuations and reproduce the turbulence statistics as well as spatial and temporal spectra with commendable accuracy as compared with the DNS results. The proposed model also exhibits a reasonable accuracy for predicting velocity fields at Reynolds numbers that are not used in the training process. With the aid of transfer learning, the computational cost of the proposed model is considered to be effectively low. Furthermore, applying the generated turbulent inflow conditions to an inflow-outflow simulation reveals a negligible development distance for the TBL to reach the target statistics. The results demonstrate, for the first time that transformer-based models can be efficient in predicting the dynamics of turbulent flows. It also shows that combining these models with generative adversarial networks-based models can be useful in tackling various turbulence-related problems, including the development of efficient synthetic-turbulent inflow generators.

\end{abstract}
%\begin{keywords}
%\end{keywords}

\section{Introduction}\label{sec:introduction}
The generation of turbulent inflow conditions is essential in simulating spatially-developing turbulent boundary layers (TBLs), considering its effect on the accuracy of the simulations and computational cost. It is also a challenging topic due to the need for the time-dependent turbulent inflow data to be accurately described. The generated data should satisfy the momentum and continuity equations and consequently match the turbulent statistics and spectra of the flow. Several approaches have been proposed to generate turbulent inflow conditions with different levels of success \citep{Wu2017}. Adding infinitesimal perturbations on the laminar mean velocity profile at the inlet section of the computational domain and allowing the transition of the boundary layer is a straightforward approach that guarantees a realistic spatially-developing TBL. However, the need for a development distance that is long enough for the flow to reach the fully-turbulent state can result in a high computational cost, making this approach not applicable for most turbulence simulations, where fully-turbulent inflow conditions are required. The use of precursor (auxiliary) parallel flow (fully developed flow) simulations with periodic boundary conditions applied to the streamwise direction is another approach that can be used by extracting flow fields from a plane normal to the streamwise direction and applying the data as inflow conditions to the main simulations. Although this method can produce accurate turbulence statistics and spectra for fully developed flows, it requires a high computational cost. Additionally, the streamwise periodicity effect, caused by the recycling of the flow within a limited domain size, can lead to physically unrealistic streamwise-repetitive features in the flow fields \citep{Wu2017}. Furthermore, using parallel flow data as inflow for a simulation of a spatially-developing TBL can result in a long development distance downstream of the domain inlet to produce the correct boundary layer characteristics \citep{Lund1993}. \par

To address this issue, a recycling-rescaling method was introduced by \cite{Lundetal1998}, which is a modified version of the method by \cite{Spalart1988}. Here the velocity fields in the auxiliary simulation are rescaled before being reintroduced at the inlet section. Another well-known approach for generating turbulent inflow conditions is adding random fluctuations based on known turbulence statistics. The methods that are based on this approach are usually called synthetic turbulent inflow generation methods. Several methods, such as the synthetic random Fourier method \citep{Leetal1997}, synthetic digital filtering method \citep{Kleinetal2003}, synthetic coherent eddy method \citep{Jarrinetal2006}, synthetic vortex method \citep{Sergent2020, Matheyetal2006, Yousif&Lim2021}, synthetic volume-force method \citep{Schlatter&Orlu2012, Spille-Kohoff&Kaltenbach2001} and numerical counterpart of the experimental tripping methods \citep{Sanmiguelvilaetal2017} have been proposed to feature a fast generation of turbulence with various levels of precision. However, a long-distance downstream of the domain inlet is required to allow the boundary layer to recover from the unphysical random fluctuations of the generated velocity fields and produce the right flow characteristics, resulting in a high computational cost. Another approach based on proper orthogonal decomposition (POD) and Galerkin projection has been proposed to build a reduced-order flow model and generate turbulent inflow conditions by utilising the most energetic eddies \citep{Johansson&Andersson2004}. A similar approach has been applied to experimental measurements \citep{Druaultetal2004, Perretetal2008} to reconstruct turbulent inflow velocity fields from hot-wire anemometry and particle image velocimetry using POD and linear stochastic estimation. This approach showed the possibility of utilising the experimental results as turbulent inflow conditions. However, the costly experimental setup makes this approach not applicable as a general method to generate turbulence. \par

The rapid development of deep learning algorithms and the increase in the graphic processing unit (GPU) capability, accompanied by the enormous amounts of high-fidelity data generated from experimental and numerical simulations, encourage exploring new data-driven approaches that can efficiently tackle various fluid-flow problems \citep{Bruntonetal2020, Kutz2017, Vinuesa&Brunton2021}. Deep learning is a subset of machine learning, where deep neural networks are used for classification, prediction and feature extraction \citep{LeCunetal2015}. Recently, several models have shown great potential in solving different problems in the field of turbulence, such as turbulence modelling \citep{Duraisamyetal2019, Wangetal2017}, turbulent flow prediction \citep{Lee&You2019, Srinivasanetal2019}, reduced-order modelling \citep{Nakamuraetal2021, Yousif&Lim2022}, flow control \citep{Fanetal2020, Park&Choi2020, Rabaultetal2019, Vinuesaetal2022}, non-intrusive sensing \citep{Guastonietal2021, Guemesetal2021} and turbulent flow reconstruction \citep{Dengetal2019, Eivazietal2022, Fukamietal2019a, Kimetal2021, Yousifetal2021, Yousifetal2022b}. \par

Furthermore, recent studies on the generation of turbulent inflow conditions using deep learning models (DLMs) have shown promising results. \cite{Fukamietal2019b} showed that convolutional neural networks (CNNs) could be used to generate turbulent inflow conditions by proposing a model based on a convolutional autoencoder (CAE) with a multilayer perceptron (MLP) to generate turbulent inflow conditions using turbulent channel flow data. \cite{Kim&Lee2020} proposed a generative adversarial network (GAN) and a recurrent neural network (RNN)-based model as a representative of unsupervised deep learning to generate turbulent inflow conditions at various Reynolds numbers using data of turbulent channel flow at various friction Reynolds numbers. Recently, \cite{Yousifetal2022a} utilised a combination of a multiscale CAE with a subpixel convolution layer ( MSC\textsubscript{SP}-AE ) having a physical constrains-based loss function and a long short-term memory (LSTM) model to generate turbulent inflow conditions from turbulent channel flow data.\par

In all the above models, the prediction of the turbulent inflow conditions is based on parallel flows, which, as mentioned before, is more suitable as inflow for fully developed TBLs. Therefore, it is necessary to develop a model that considers the spatial development of TBLs \citep{Jimenezetal2010}. In this context, this paper proposes a deep learning method consisting of a transformer and a multiscale-enhanced super-resolution generative adversarial network (MS-ESRGAN) to generate turbulent inflow conditions for spatially-developing TBL simulations. \par

The remainder of this paper is organised as follows. In Section \ref{sec:2}, the methodology of generating the turbulent inflow data using the proposed DLM is explained. The datasets used for training and testing the model are described in Section \ref{sec:3}. Section \ref{sec:4} presents the results obtained from testing the proposed model. Finally, Section \ref{sec:5} presents the conclusions. \par

\section{Methodology}\label{sec:2}

\begin{figure}
  \centerline{\includegraphics[scale=0.12]{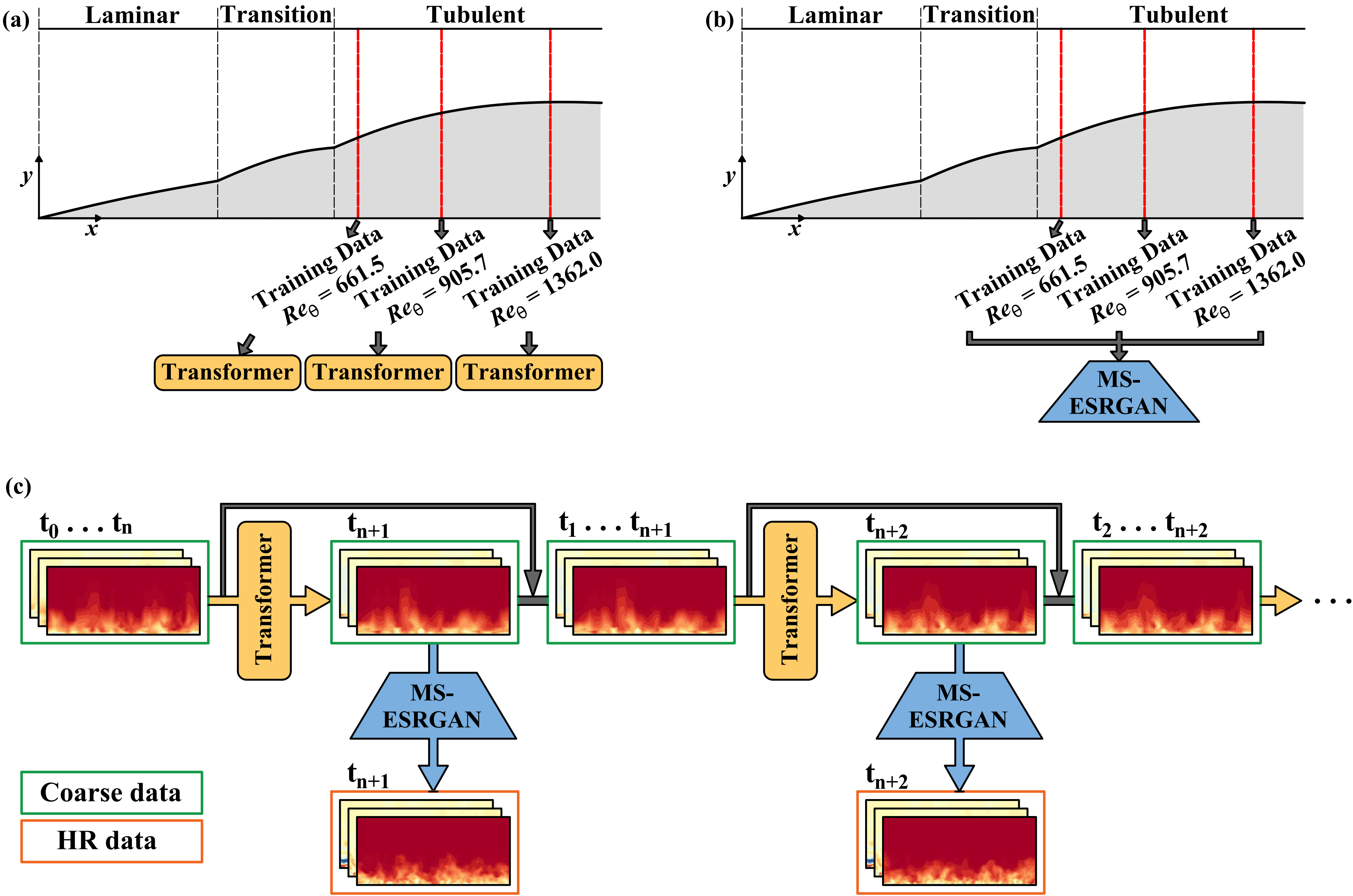}}% Images in 100% size
  \caption{Schematic of (a) training procedure for the transformer, (b) training procedure for the MS-ESRGAN and (c) turbulent inflow generation using the proposed DLM.}
\label{fig:k1}
\end{figure}

The proposed DLM is a combination of two architectures. The first one is the transformer \citep{Vaswanietal2017} and the second one is the MS-ESRGAN \citep{Yousifetal2021}. The transformer is used to predict the temporal evolution of extremely coarse velocity fields obtained by selecting distributed points at various sections along the streamwise direction of a spatially-developing TBL flow, as shown in figure~\ref{fig:k1}(a). Here the flow data are obtained through direct numerical simulation (DNS). Meanwhile, the MS-ESRGAN is used to perform a super-resolution reconstruction of the data for all the sections predicted by the transformer, leading to a final high-resolution (HR) data, i.e. velocity fields with the same resolution as the ground truth data, as shown in figure~\ref{fig:k1}(b). In other words, the transformer is trained for the data at each section, whereas MS-ESRGAN is trained for all the sections used in the training process. Figure~\ref{fig:k1}(c) shows the schematic representation of the proposed DLM for generating turbulent inflow conditions. As shown in the figure, the input to the DLM is represented by coarse velocity data obtained from a plane normal to the streamwise direction with time interval [$t_0, ... ,t_n$], and the output is represented by predicted high-resolution velocity data at instants, $t_{n+1}$, where $n$ is set to 12 in this study.\par
In this study, the open-source library TensorFlow 2.4.0 \citep{Abadietal2016} is used for the implementation of the presented model. The source code of the model is available on the following web page: \vspace{0.1cm}
 
Click the link: \href{https://fluids.pusan.ac.kr/fluids/65416/subview.do}{https://fluids.pusan.ac.kr/fluids/65416/subview.do} \par

\subsection {Transformer}\label{sec:Transformer}

LSTM \citep{Hochreiter&Schmidhuber1997} is an artificial neural network that can handle sequential data and time-series modelling. LSTM is a type of RNN \citep{Rumelhartetal1986}. It has also played an essential role in modelling the temporal evolution of turbulence in various problems \citep{Eivazietal2021, Kim&Lee2020, Nakamuraetal2021,Srinivasanetal2019, Yousif&Lim2022}. Although LSTM is designed to overcome most of the traditional RNN limitations, such as vanishing gradients and explosion of gradients \citep{Graves2012}, it is usually slow in terms of training due to its architecture, which requires that the time-series data are introduced to the network sequentially. This prevents parallelisation of the training process, which is why GPU is used in deep learning calculations. Furthermore, LSTM has shown a limitation in dealing with long-range dependencies. \par

The transformer \citep{Vaswanietal2017} was introduced to deal with these limitations by applying the self-attention concept to compute the representations of its input and output data without feeding the data sequentially. In this study, a transformer is used to model the temporal evolution of the velocity fields that represent the turbulent inflow data. \par

\begin{figure}
  \centerline{\includegraphics[scale=0.12]{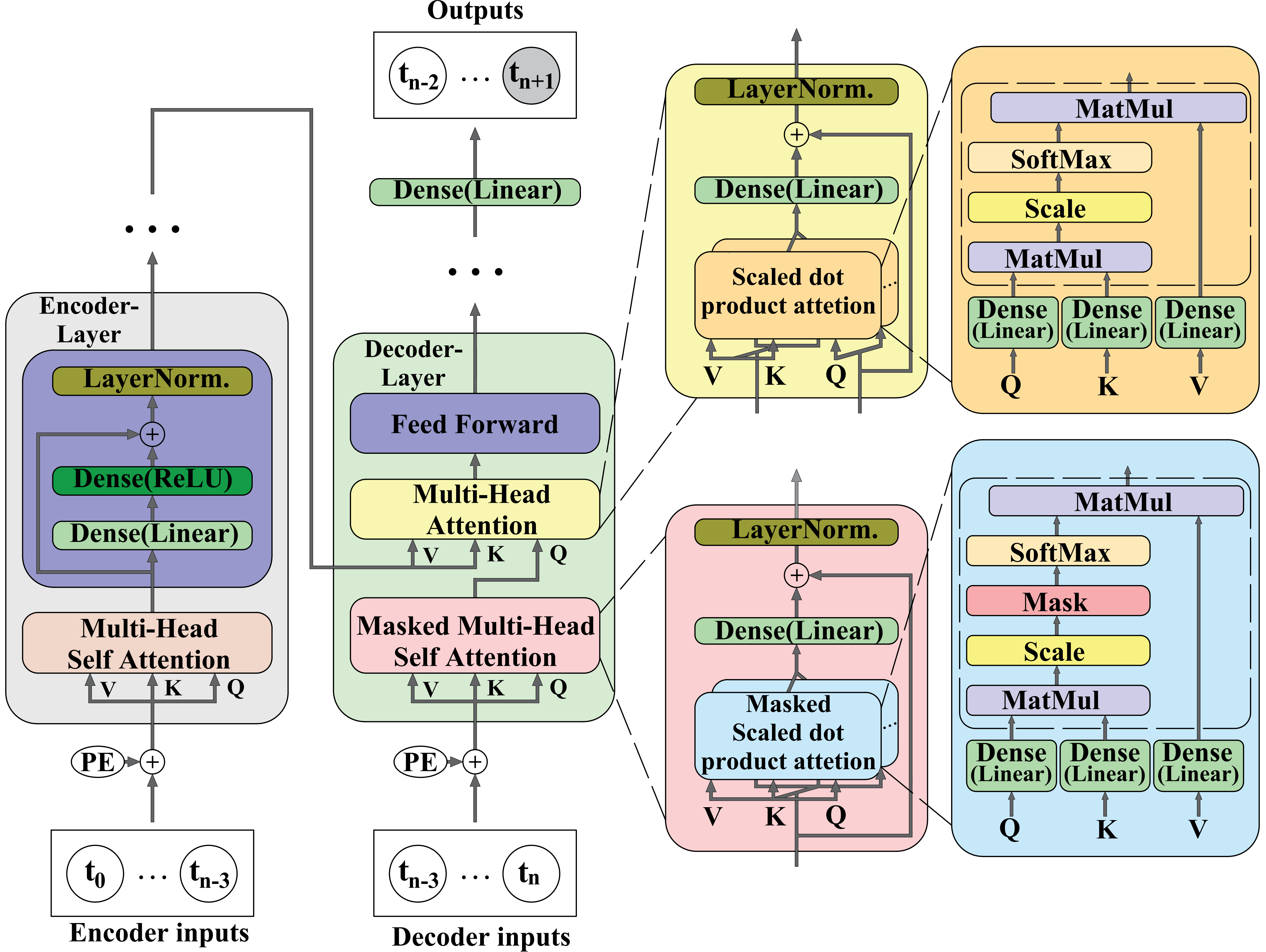}}% Images in 100% size
  \caption{Architecture of the transformer. The dashed line box represents the scaled dot-product attention.}
\label{fig:k2}
\end{figure}

Figure~\ref{fig:k2} shows the transformer used in this study. Similar to the original transformer proposed by \cite{Vaswanietal2017}, it has two main components: encoder and decoder. The inputs of both components are passed through a positional encoding using sine and cosine functions, which can encode the order information of the input data into a vector and add it directly to the input vector. The encoder consists of six stacked encoder layers. Each layer contains a multi-head self-attention sublayer and a feed-forward sublayer. The input of the multi-head self-attention sublayer consists of queries ($Q$), keys ($K$), and values ($V$). Note that attention is a function that can map a query and set of key-value pairs to output, where the queries, keys, values and output are all vectors. The output can be calculated as the weighted sum of the values. The attention function is represented by scaled dot-product attention, which is an attention mechanism with the dot products are scaled down by $\sqrt{d_k }$, where $d_k$ is the dimension of $Q$, $K$ and it is equal to the dimension of $V$, i.e. $d_v$. \par

The scaled dot-product attention is calculated as follows:

\begin{equation} \label{eqn:eq1}
\mathrm{Attention} (Q, K, V) = \mathrm{softmax} \left( \frac{Q K^T}{\sqrt{d_k}} \right) V ,
\end{equation}

\noindent where softmax is a function that takes an input vector and normalizes it to a probability distribution so that the output vector has values that sum to 1 \citep{Goodfellowetal2014}.
In the multi-head self-attention sublayer, $d_v = d_{model}/h$, where $d_{model}$ and $h$ are the dimension of the input data to the model and number of heads, respectively. \par

The multi-head attention allows the model to jointly attend to information from different representation subspaces at different positions such that

\begin{equation} \label{eqn:eq2}
\mathrm{Multi-Head} (Q, K, V) = Concat \left( head_1 , ..., head_h \right) W^o ,
\end{equation}

\begin{equation} \label{eqn:eq3}
head_i = \mathrm{Attention} \left( Q W_i^Q , K W_i^K , V W_i^V \right), 
\end{equation}

\noindent where $W_i^Q$, $W_i^K$ and $W_i^V$ are the weights corresponding to $Q$, $K$, $V$ at every head, respectively; $W^o$ represents the weights of the concatenated heads. $W_i^Q \in \mathbb{R}^{d_{model} \times d_k }$, $W_i^K \in \mathbb{R}^{d_{model} \times d_k }$, $W_i^V \in \mathbb{R}^{d_{model} \times d_v }$ and $W^o \in \mathbb{R}^{ hd_v \times d_{model} }$.  \par
The multi-head self-attention sublayer contains six heads of scaled dot-product attention.\par
A residual connection \citep{Heetal2016} is applied around the multi-head attention, followed by layer normalisation \citep{Baetal2016}. \par

The second part of the encoder layer, i.e. the feed-forward sublayer, contains two dense layers with linear and rectified linear unit (ReLU) activation functions. This layer projects the vector to a larger space, where it is easier to extract the required information and then projects it back to the original space. As in the multi-head self-attention sublayer, a residual connection is employed before applying layer normalisation.

Similar to the encoder, the decoder contains six decoder layers. In addition to the multi-head self-attention and feed-forward sublayers, the decoder layer has a third sublayer that performs multi-head attention over the output of the encoder stack. Furthermore, the multi-head self-attention sublayer is changed to a masked multi-head self-attention sublayer, as shown in figure~\ref{fig:k2}, which is similar to the multi-head self-attention sublayer with the difference that the scaled dot-product attention is changed to a masked scaled dot-product attention \citep{Vaswanietal2017}. The masking operation ensures that the prediction can only depend on the known outputs, a fact that prevents later information leakage. In this study, the dropout technique is applied to every sublayer before the residual connection and the rate of dropout is set to 0.1. \par

The square of the $L_2$ norm error is chosen as a loss function for the transformer such that

\begin{equation} \label{eqn:eq4}
\mathcal{L}_{transformer}= \frac{1}{M} \sum_{m=1}^M \left\| {Output}_m - {Target}_m \right\|_2^2            
\end{equation}

\noindent where $Output$ and $Target$ represent the output from the transformer and ground truth data, respectively, at a specific time step, $m$. $M$ represents the size of the training mini-batch, which is set to 64.
The adaptive moment estimation (Adam) optimisation algorithm \citep{Kingma&Ba2017} is used to update the weights of the model.

\subsection {MS-ESRGAN}\label{sec:MS-ESRGAN}

GANs \citep{Goodfellowetal2014} have shown great success in image transformation and super-resolution problems \citep{Ledigetal2017, Mirza&Osindero2014, Wangetal2018, Zhuetal2017}. GAN-based models have also shown promising results in reconstructing HR turbulent flow fields from coarse data \citep{Fukamietal2021, Fukamietal2019a, Guemesetal2021, Kimetal2021,  Yousifetal2021, Yousifetal2022b}. In a GAN model that is used for image generation, two adversarial neural networks called the generator ($G$) and the discriminator ($D$) compete with each other. $G$ tries to generate artificial images with the same statistical properties as those of the real ones, whereas $D$ tries to distinguish the artificial images from the real ones. After successful training, $G$ should be able to generate artificial images that are difficult to distinguish by $D$. This process can be expressed as a min-max two-player game with a value function $V(D, G)$ such that

\begin{equation} \label{eqn:eq5}
\begin{split}
\substack{min\\G} ~\substack{max\\D} ~V(D,G) = \mathbb{E}_{x_r \sim P_{data}(x_r)} [ {\rm log} D(x_r )] + \mathbb{E}_{\xi \sim P_\xi(\xi) } [ {\rm log} (1-D(G(\xi)))],
\end{split}
\end{equation}

\noindent where $x_r$ is the image from the ground truth data (real image) and $P_{data}(x_r )$ is the real image distribution. $\mathbb{E}$ represents the operation of calculating the average of all the data in the training mini-batch. In the second right term of equation~\ref{eqn:eq5}, $\xi$ is a random vector used as an input to $G$ and $D(x_r)$ represents the probability that the image is real and not generated by the generator. The output from $G$, i.e. $G(\xi)$, is expected to generate an image that is similar to the real one, such that the value of $D(G(\xi))$ is close to 1. Meanwhile, $D(x_r)$ returns a value close to 1, whereas $D(G(\xi))$ returns a value close to 0. Thus, in the training process, $G$ is trained in a direction that minimises $V(D,G)$, whereas $D$ is trained in a direction that maximises $V(D,G)$.

\begin{figure}
  \centerline{\includegraphics[scale=0.12]{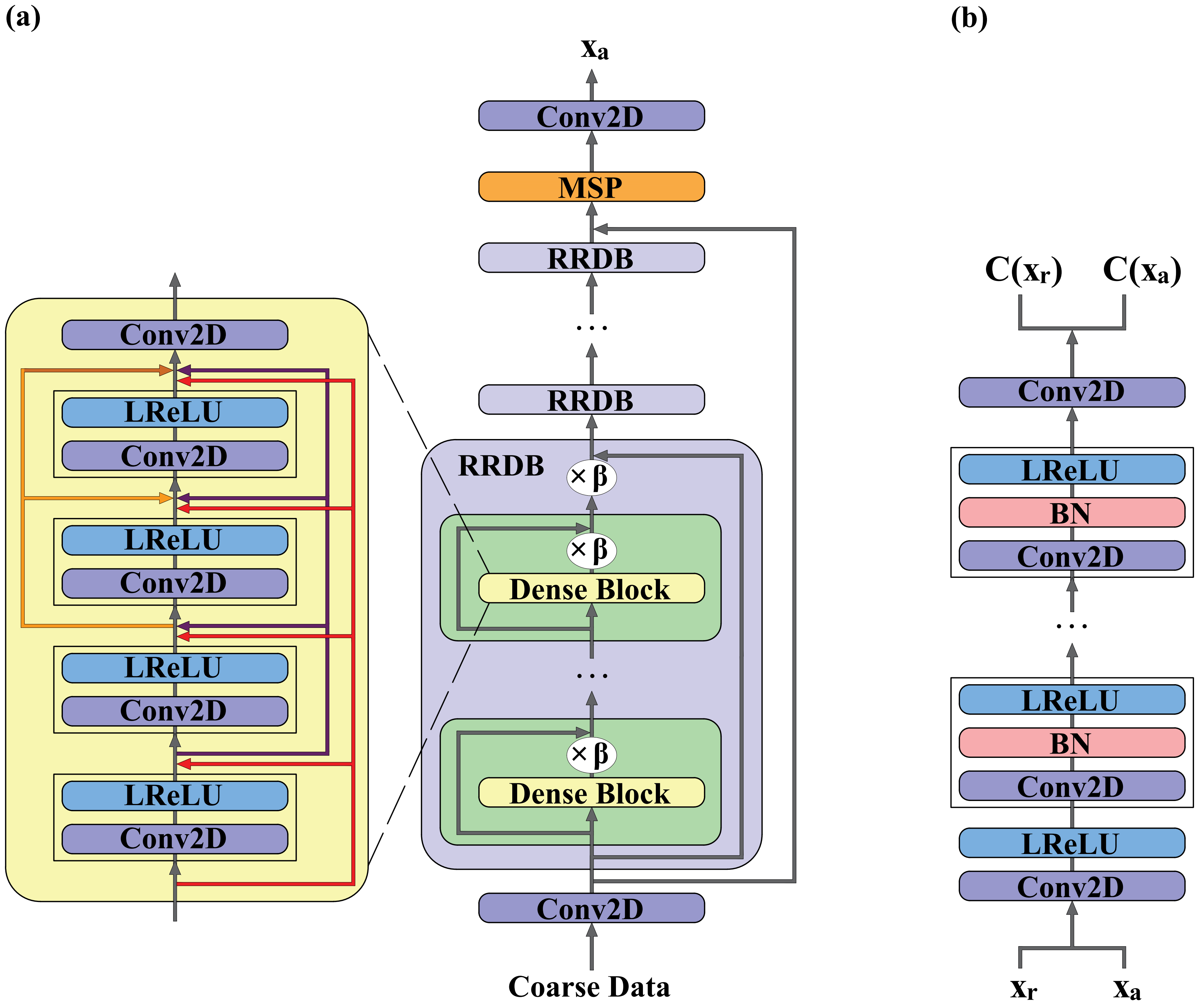}}% Images in 100% size
  \caption{Architecture of MS-ESRGAN. (a) The generator, where $\beta$ is the residual scaling parameter which is set to 0.2 in this study, and (b) the discriminator.}
\label{fig:k3}
\end{figure}

In this study, MS-ESRGAN \citep{ Yousifetal2021} is used to perform super-resolution reconstruction of the velocity fields predicted by the transformer. MS-ESRGAN is based on the enhanced super-resolution GAN (ESRGAN) \citep{Wangetal2018}. Figure~\ref{fig:k3} shows the architecture of MS-ESRGAN. As shown in figure~\ref{fig:k3}(a), $G$ consists of a deep CNN represented by residual in residual dense blocks (RRDBs) and multiscale parts (MSP). Note that the input to $G$ is low-resolution data, which are first passed through a convolutional layer and then through a series of RRDBs. The MSP, consisting of three parallel convolutional sub-models with different kernel sizes, is applied to the data features extracted by RRDBs. More details for MSP can be found in \cite{Yousifetal2021, Yousifetal2022b}. The outputs of the three sub-models are summed and passed through a final convolutional layer to generate HR artificial data $(x_a)$. Figure~\ref{fig:k3}(b) shows that the artificial and real data are fed to $D$ and passed through a series of convolutional, batch normalisation and leaky ReLU layers. As a final step, the data are crossed over a convolutional layer. The non-transformed discriminator outputs using the real and artificial data, i.e. $C(x_r )$ and $C(x_a )$, are used to calculate the relativistic average discriminator value $D_{Ra}$ \citep{Jolicoeur-Martineau2018}:

\begin{equation} \label{eqn:eq6}
D_{Ra} (x_r , x_a ) = \sigma (C \left(x_r ) \right) - \mathbb{E}_{x_a} \left[C ( x_a) \right],
\end{equation}

\begin{equation} \label{eqn:eq7}
D_{Ra} (x_a , x_r ) = \sigma (C \left(x_a) \right) - \mathbb{E}_{x_r} \left[C ( x_r) \right],
\end{equation}

\noindent where $\sigma$ is the sigmoid function. In equations~\ref{eqn:eq6} and \ref{eqn:eq7}, $D_{Ra}$ represents the probability that the output from $D$ using the real image is relatively more realistic than the output using the artificial image. \par

Then, the discriminator loss function is defined as follows:

\begin{equation} \label{eqn:eq8}
\ell_D^{Ra} = -\mathbb{E}_{x_r} \left[ {\rm log} (D_{Ra} (x_r , x_a)) \right] - \mathbb{E}_{x_a} \left[ {\rm log} (1 - D_{Ra} (x_a, x_r )) \right].
\end{equation}

The adversarial loss function of the generator can be expressed in a symmetrical form as follows:

\begin{equation} \label{eqn:eq9}
\ell_G^{Ra} = -\mathbb{E}_{x_r} \left[ {\rm log} (1 - D_{Ra} (x_r , x_a )) \right] - \mathbb{E}_{x_a} \left[ {\rm log} (D_{Ra} (x_a , x_r )) \right].
\end{equation}

The total loss function of the generator is defined as

\begin{equation} \label{eqn:eq10}
\mathcal{L}_G =  \ell_G^{Ra} + \beta \ell_{pixel} + \ell_{perceptual},     
\end{equation}

\noindent where $\ell_{pixel}$ is the error calculated based on the pixel difference of the generated and ground truth data; $\ell_{perceptual}$ represents the difference between features that are extracted from the real and the artificial data. The pre-trained CNN VGG-19 \citep{Simonyan&Zisserman2015} is used to extract the features using the output of three different layers \citep{Yousifetal2021}. Here, $\beta$ is a weight coefficient and its value is set to be 5000. The square of the $L_2$ norm error is used to calculate $\ell_{pixel}$ and $\ell_{perceptual}$. The size of the mini-batch is set to 32. As in the transformer model, the Adam optimisation algorithm is used to update the weights of the model.\par

\section{Data description and pre-processing }\label{sec:3}

The transitional boundary layer database \citep{Lee&Zaki2018} available at the Johns Hopkins turbulence databases (JHTDB) is considered in this study for the training and testing of the DLM. The database was obtained via DNS of incompressible flow over a flat plate with an elliptical leading edge. In the simulation, the half-thickness of the plate $L$ and the free stream velocity $U_\infty$ are used as a reference length and velocity. Additionally, $x$, $y$ and $z$ are defined as the streamwise, wall-normal and spanwise coordinates, respectively, with the corresponding velocity components $u$, $v$, and $w$. Note that the same definitions of the coordinates and velocity components are used in this study. \par

In the simulation, the length of the plate, $L_x = 1,050L$ measured from the leading edge ($x$ = 0), the domain height,  $L_y = 40L$ and the width of the plate, $L_z = 240L$. The stored database in JHTDB is in the range $x \in [30.2185, 1000.065]L$, $y \in \left[0.0036, 26.488 \right]L$, and $z \in \left[ 0, 240\right]L$. The corresponding number of grid points is $N_x \times N_y \times N_z = 3320 \times 224 \times 2048 \approx 1.5 \times 10^9$. The database time step is $\Delta t = 0.25L/U_\infty$. The stored database spans the following range in momentum-thickness-based Reynolds number, $ Re_\theta = U_\infty \theta/\nu \in \left[105.5,1502.0 \right]$, where $\theta$ represents the momentum thickness and $\nu$ is the kinematic viscosity. More details for the simulation and database can be found in \cite{Lee&Zaki2018} and on the website of JHTDB.\par

In this study, the datasets within the range of $Re_\theta \in \left[ 661.5, 1502.0 \right]$ are considered for training and testing the DLM. This range of $Re_\theta$ in the database represents the fully-turbulent part of the flow \citep{Lee&Zaki2018}. Datasets of the velocity components are collected from various ($y-z$) planes along the streamwise direction, with the number of snapshots = 4700 for every plane. To reduce the computational cost, the original size of each plane, $N_y \times N_z = 224 \times 2048$ is reduced to $112 \times 1024$. Furthermore, to increase the amount of training and testing data, every selected plane is divided into four identical sections along the spanwise direction, resulting in $N_y \times N_z = 112 \times 256$ for each section. To obtain the coarse data, the size of the data is further reduced to $N_y \times N_z = 14 \times 32$, which is obtained by selecting distributed points in the fields. The distribution of the points is obtained in a stretching manner such that more points can be selected near the wall. A time series of 4000 snapshots for each section are used to train the DLM, resulting in a total number of training snapshots = $ 4000\times4\times3=48000$. The fluctuations of the velocity fields are used in the training and prediction processes. The input data to the DLM are normalised using the min-max normalisation function to produce values between 0 and 1.\par

\section{Results and discussion} \label{sec:4}

\subsection{Results from the DLM trained at various $Re_\theta$}

\begin{figure}
  \centerline{\includegraphics[scale=0.12]{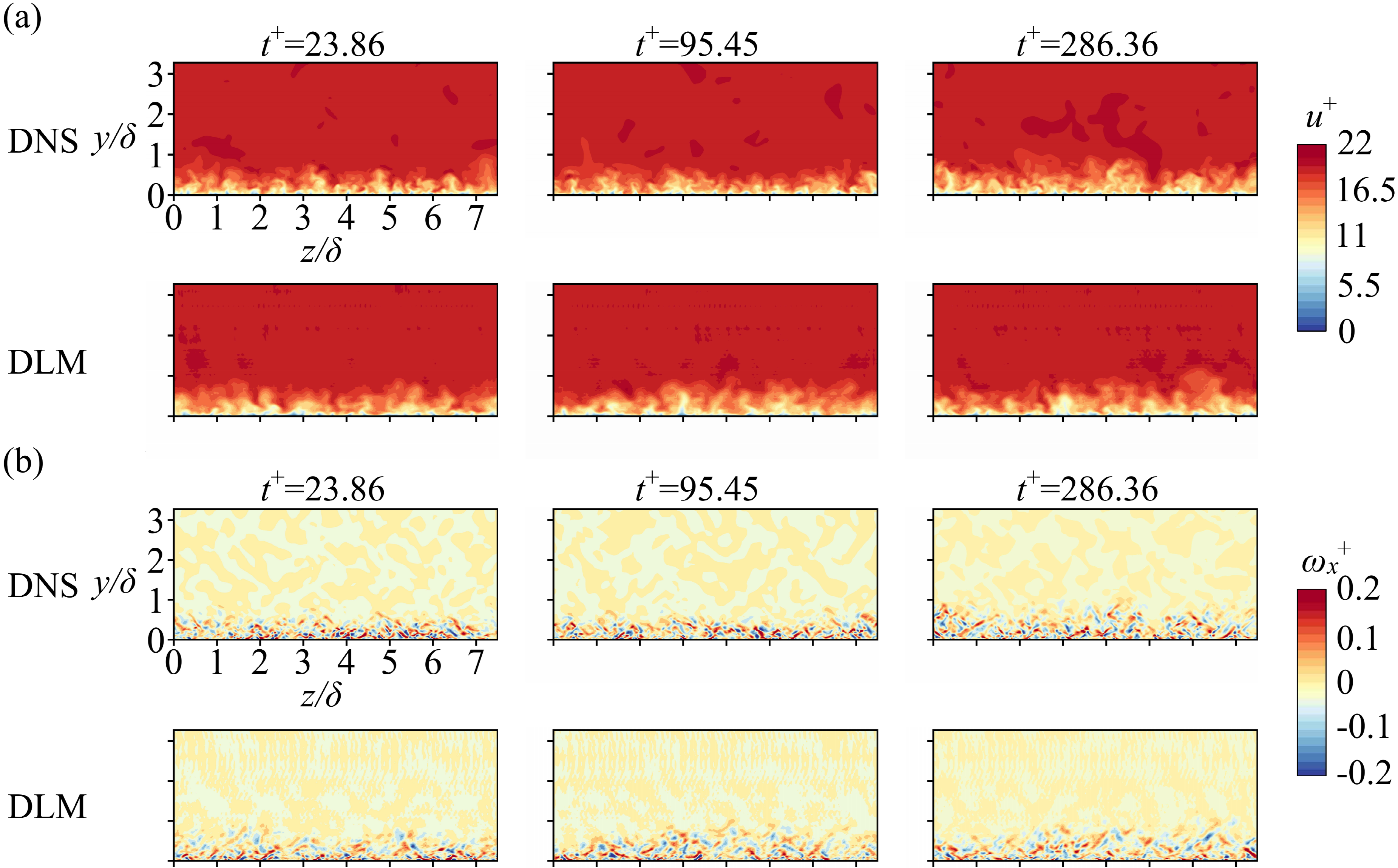}}% Images in 100% size
  \caption{Instantaneous (a) streamwise velocity and (b) vorticity fields at $Re_\theta$ = 661.5, for three different instants. Reference (DNS) and predicted (DLM) data are shown.}
\label{fig:k4}
\end{figure}

\begin{figure}
  \centerline{\includegraphics[scale=0.12]{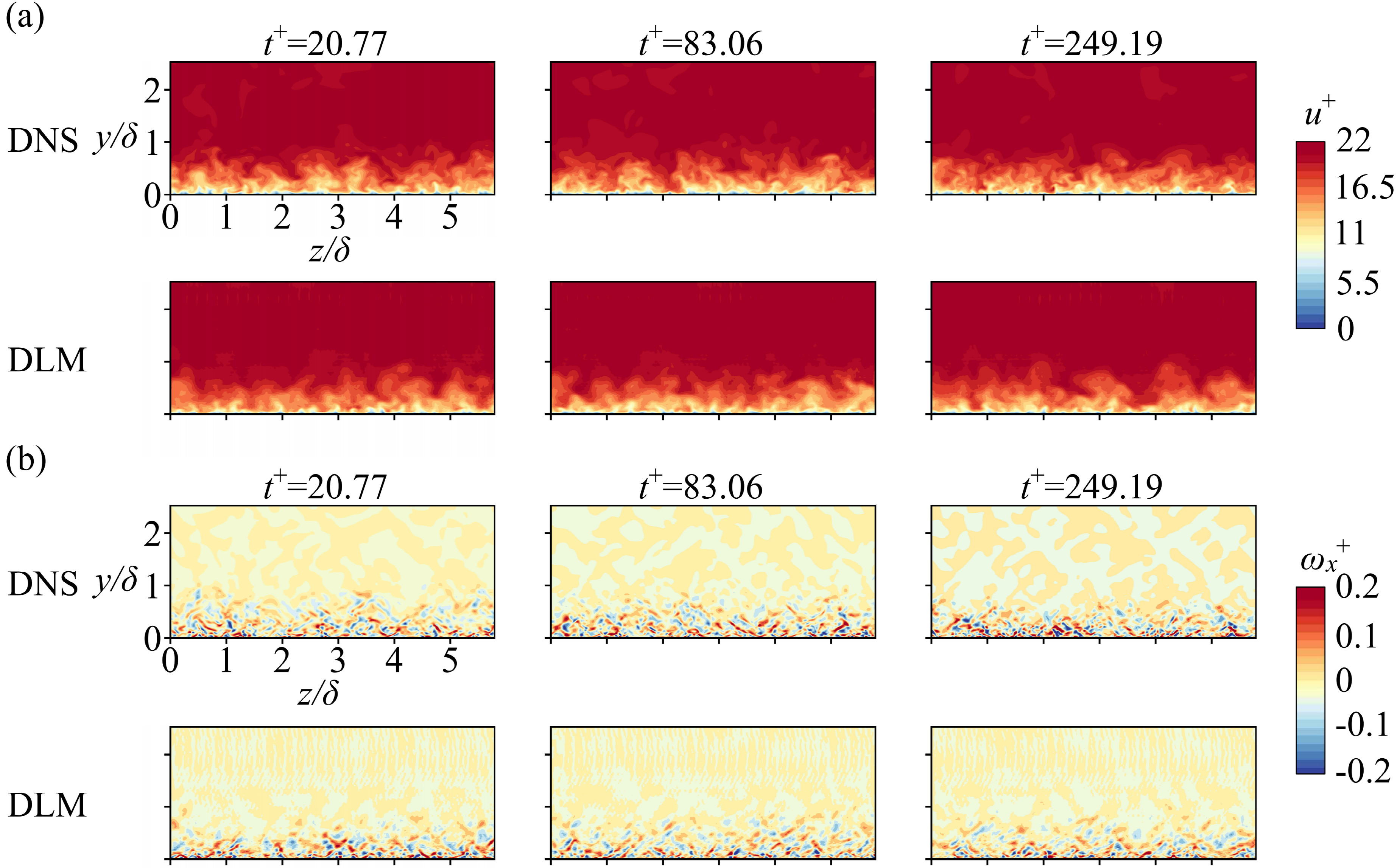}}% Images in 100% size
  \caption{Instantaneous (a) streamwise velocity and (b) vorticity fields at $Re_\theta$ = 905.7, for three different instants. Reference (DNS) and predicted (DLM) data are shown.}
\label{fig:k5}
\end{figure}

\begin{figure}
  \centerline{\includegraphics[scale=0.12]{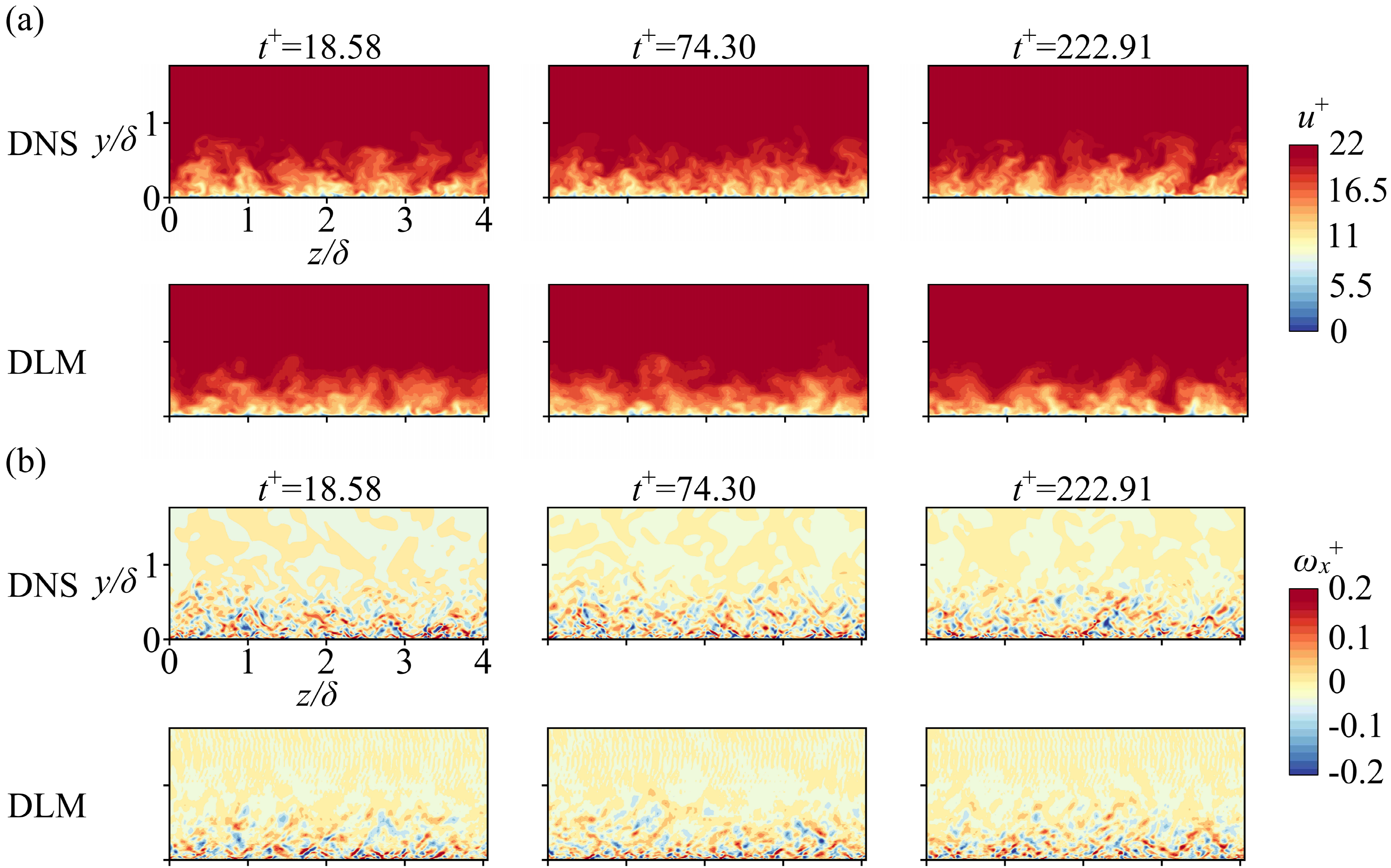}}% Images in 100% size
  \caption{ Instantaneous (a) streamwise velocity and (b) vorticity fields at $Re_\theta$ = 1362.0, for three different instants. Reference (DNS) and predicted (DLM) data are shown.}
\label{fig:k6}
\end{figure}

This section examines the capability of the proposed DLM to generate turbulent inflow data at three different Reynolds numbers for which the network has already been trained, $Re_\theta$ = 661.5, 905.7, and 1362.0. Figures~\ref{fig:k4}-\ref{fig:k6} show the instantaneous streamwise velocity ($u^+$) and vorticity ($\omega_x^+$) fields of the DNS and the predicted data for three different time steps, where the superscript ‘$+$’ denotes normalisation by viscous inner scale; in the figures, $\delta$ represents the boundary layer thickness. The figures show that the instantaneous flow fields can be predicted using the model with a commendable agreement with the DNS data. Note that the model has shown a capacity to predict the instantaneous flow fields for a long period of time, more than the one required for the flow data to reach a statistically stationary state (reaching fixed first and second-order statistics over time), i.e. for a number of time steps = 10000.\par

\begin{table}
  \begin{center}
\scalebox{1.0}{
\begin{tabular}{cccccc}
&$Re_\theta$~ ~&~~$661.5$~ ~&~~$905.7$~ ~&~~$1362.0$~ ~& \\\\
&$\mathrm {DNS}$~ ~&~~$1.458$~~&~~$1.457$~~&~~$1.453$~~&   \\
&$\mathrm {DLM}$~ ~&~~$1.415$~~&~~$1.423$~~&~~$1.429$~~& \\  
    \end{tabular}}
  \caption{Shape factor values of the DNS and predicted planes by the DLM.}
  \label{tab:Table1}
  \end{center}
\end{table}

The shape factor (ratio of displacement to momentum thickness) values of the DNS and predicted planes are shown in table~\ref{tab:Table1}. A slight under-prediction can be seen in all the predicted values with the highest deviation of 2.9$5\%$ at  $Re_\theta$ = 661.5.

\begin{figure}
  \centerline{\includegraphics[scale=0.12]{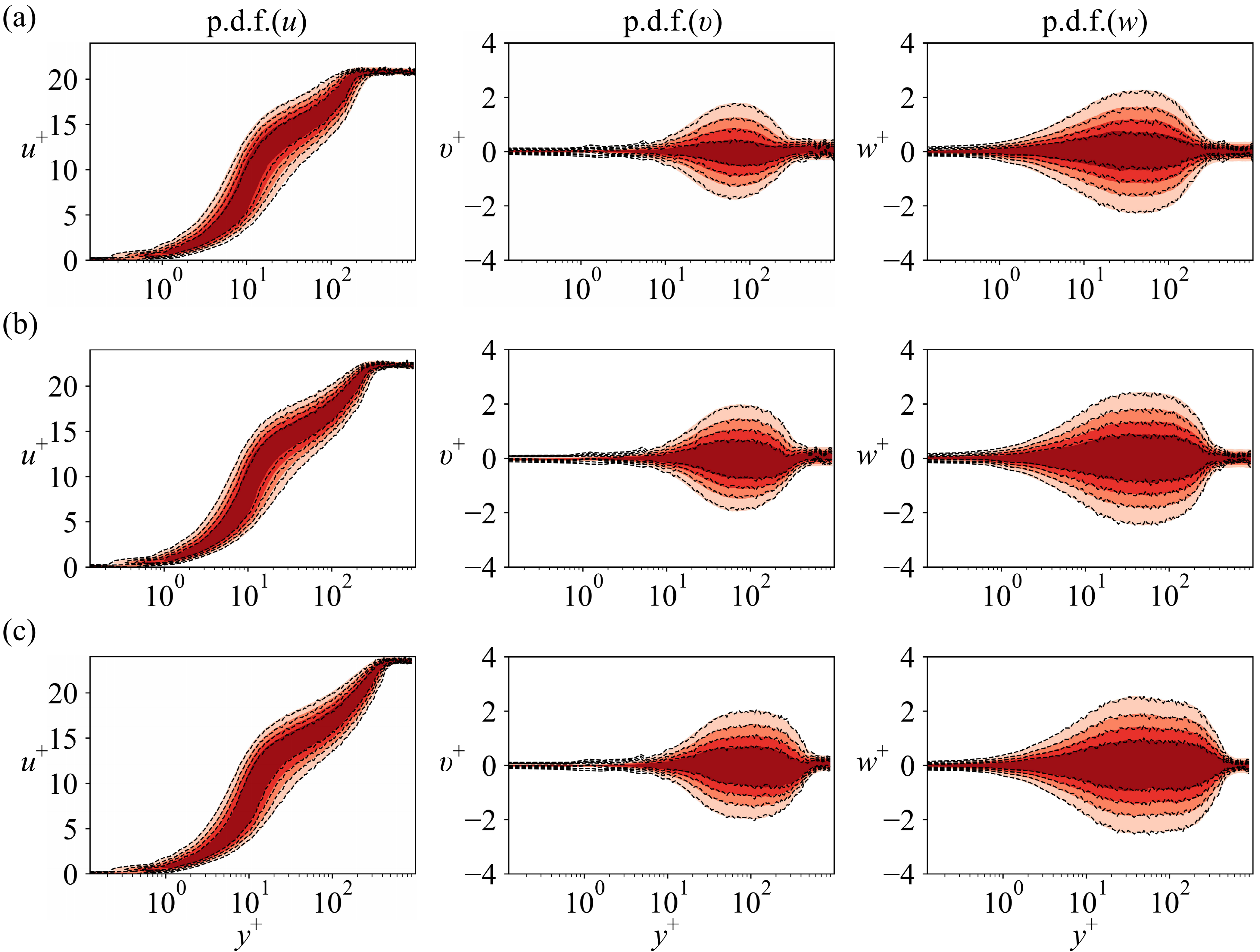}}% Images in 100% size
  \caption{Probability density functions of the velocity components as a function of the wall-normal distance. The shaded contours represent the results from the DNS data and the dashed ones represent the results from the predicted data. The contour levels are in the range of 20$\%$-80$\%$ of the maximum p.d.f. with an increment of 20$\%$.
(a) $Re_\theta$ = 661.5, (b) $Re_\theta$ = 905.7, (c) $Re_\theta$ = 1362.0.}
\label{fig:k7}
\end{figure}

Figure~\ref{fig:k7} shows the probability density functions (p.d.f.) of the velocity components ($u^+$, $v^+$, and $w^+$) plotted against the wall-normal distance ($y^+$). The figure shows that the p.d.f. plots of the generated velocity components are generally consistent with the p.d.f. plots obtained from the DNS data, indicating the capability of the model in predicting the velocity fields with distributions of the velocity components that are consistent with those of the DNS data.\par

\begin{figure}
  \centerline{\includegraphics[scale=0.12]{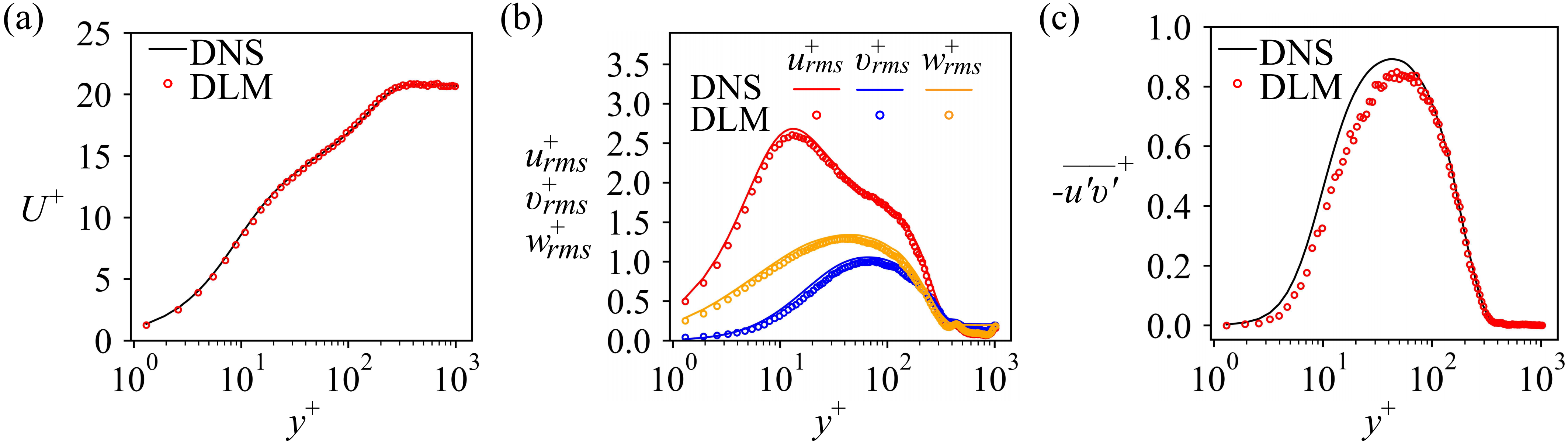}}% Images in 100% size
  \caption{Turbulence statistics of the flow at $Re_\theta$ = 661.5. (a) Mean streamwise velocity profile. (b) Root-mean-square profiles of the velocity components. (c) Reynolds shear stress profile.}
\label{fig:k8}
\end{figure}

\begin{figure}
  \centerline{\includegraphics[scale=0.12]{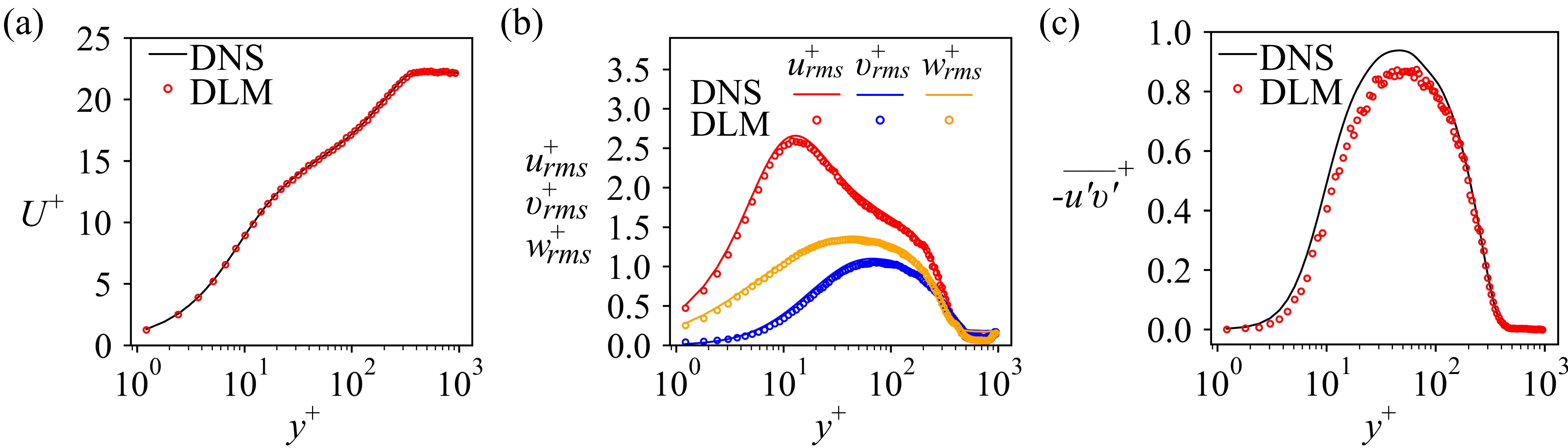}}% Images in 100% size
  \caption{Turbulence statistics of the flow at $Re_\theta$ = 905.7. (a) Mean streamwise velocity profile. (b) Root-mean-square profiles of the velocity components. (c) Reynolds shear stress profile.}
\label{fig:k9}
\end{figure}

\begin{figure}
  \centerline{\includegraphics[scale=0.12]{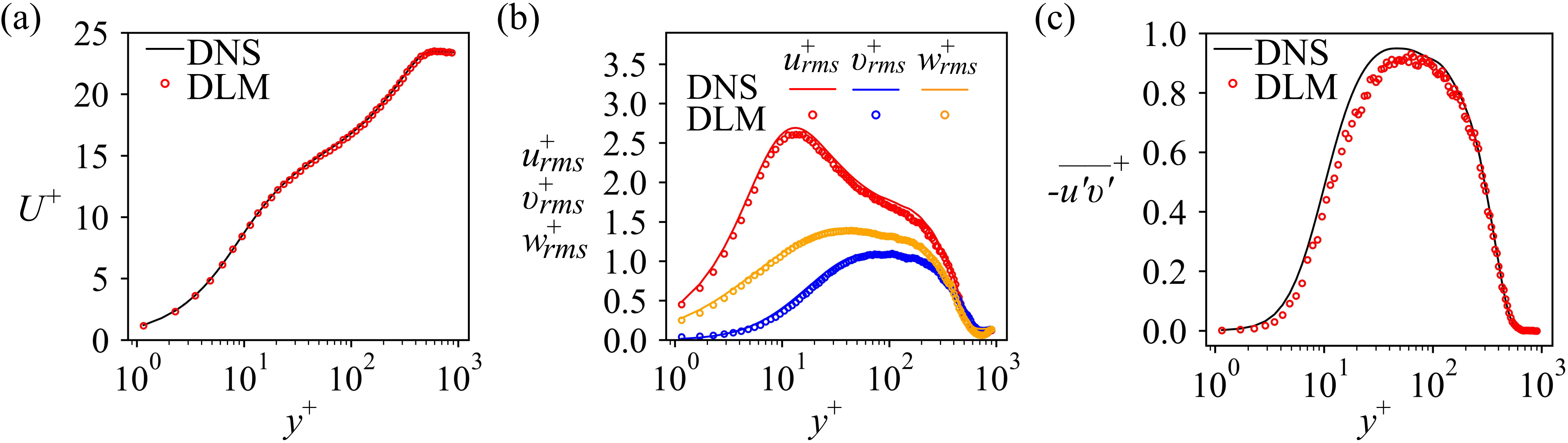}}% Images in 100% size
  \caption{Turbulence statistics of the flow at $Re_\theta$ = 1362.0. (a) Mean streamwise velocity profile. (b) Root-mean-square profiles of the velocity components. (c) Reynolds shear stress profile.}
\label{fig:k10}
\end{figure}

Figures~\ref{fig:k8}-\ref{fig:k10} compare the turbulence statistics of the generated velocity fields with the turbulence statistics of the DNS data. As shown in the figures, the mean streamwise velocity profile ($U^+$) for all the three Reynolds numbers shows excellent agreement with the results obtained from the DNS. The comparison of root-mean-square profiles of the velocity components ($u_{rms}^+$, $v_{rms}^+$, and $w_{rms}^+$) reveals good agreement with the results obtained from the DNS. However, the profile of the Reynolds shear stress ($\overline{u'v'}^+$) shows a slight under-prediction in the region between near the wall and the maximum Reynolds shear stress and the profile values in this region improve as the Reynolds number increases. This might be attributed to the fact that with the increase in the boundary-layer thickness, the effect of zero padding in the convolution processes is decreased in MS-ESRGAN, resulting in a better prediction of the velocity fields in this region of the boundary layer. These results are consistent with the results from  table~\ref{tab:Table1}. \par

\begin{figure}
  \centerline{\includegraphics[scale=0.15]{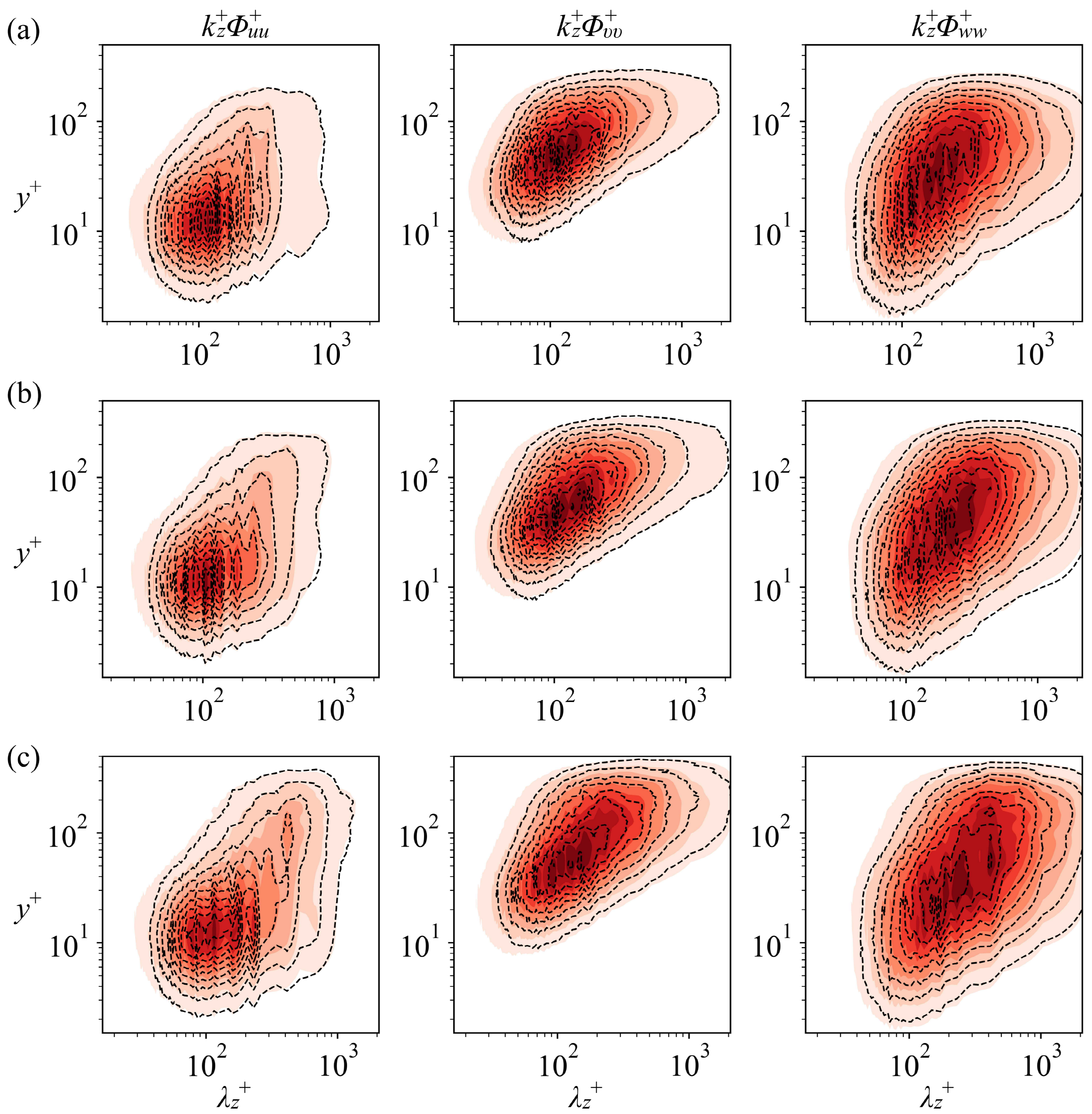}}% Images in 100% size
  \caption{Premultiplied spanwise wavenumber energy spectra of the velocity components as a function of the wall-normal distance and the spanwise wavelength. The shaded contours represent the results from the DNS data and the dashed ones represent the results from the predicted data. The contour levels are in the range of 10$\%$\textendash 90$\%$ of the maximum $k_z^+ \Phi_{\alpha \alpha}^+$ with an increment of 10$\%$. (a) $Re_\theta$ = 661.5, (b) $Re_\theta$ = 905.7, (c) $Re_\theta$ = 1362.0.}
\label{fig:k11}
\end{figure}

The capability of the proposed DLM to produce realistic spatial spectra of the velocity fields is investigated by employing the premultiplied spanwise wavenumber spectra, $k_z \Phi_{\alpha \alpha}$, where $\Phi_{\alpha \alpha}$ represents the spanwise wavenumber spectra, $\alpha$ represents the velocity component and $k_z$ is the spanwise wavenumber. Figure~\ref{fig:k11} shows the contour plots of $k_z^+ \Phi_{\alpha \alpha}^+$ as a function of $y^+$ and the spanwise wavelength, $\lambda_z^+$. The figure shows that the spectra of the velocity components are generally consistent with those obtained from the DNS data with a slight deviation at the high wavenumbers. This indicates that the two-point correlations of the generated velocity components are consistent with those obtained from the DNS data, further supporting the excellent performance of the proposed DLM to properly represent the spatial distribution of the velocity fields. It is worth noting that the ability of the model to reproduce accurate spectra is essential in generating the turbulent inflow conditions to guarantee that the turbulence will be sustained after introducing the synthetic-inflow; otherwise, the generated inflow would require very long distances to reach "well-behaved" turbulent conditions, and these fluctuations could also dissipate. \par

\begin{figure}
  \centerline{\includegraphics[scale=0.15]{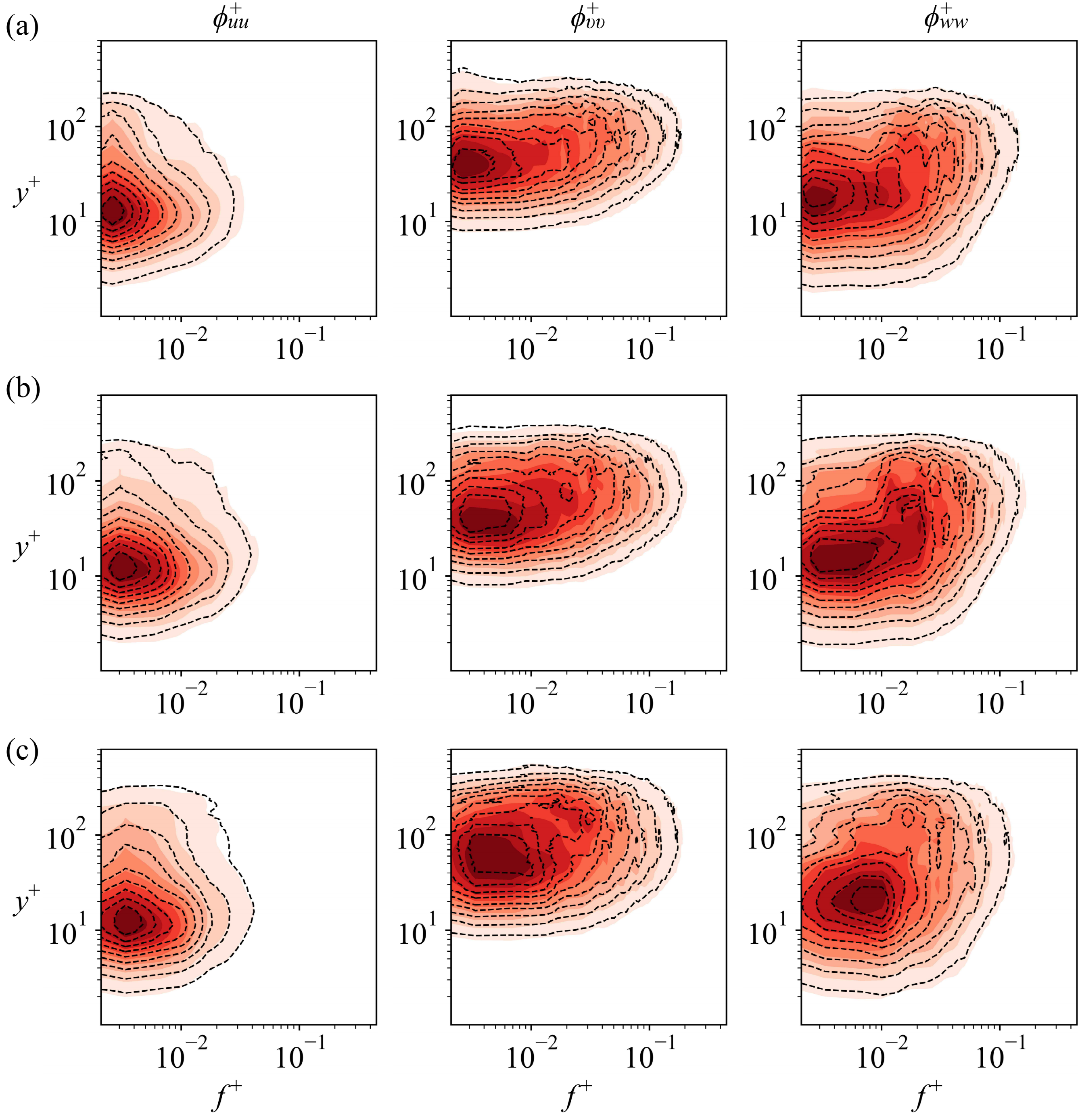}}% Images in 100% size
  \caption{Frequency spectra of the velocity components as a function of the wall-normal distance and the frequency. The shaded contours represent the results from the DNS data and the dashed ones represent the results from the predicted data. The contour levels are in the range of 10$\%$\textendash 90$\%$ of the maximum $ \phi_{\alpha \alpha}^+$ with an increment of 10$\%$. (a) $Re_\theta$ = 661.5, (b) $Re_\theta$ = 905.7, (c) $Re_\theta$ = 1362.0.}
\label{fig:k12}
\end{figure}

To evaluate the performance of the proposed DLM to generate the velocity fields with accurate dynamics, the frequency spectra, $\phi_{\alpha \alpha}^+$, as a function of $y^+$ and the frequency, $f^+$ are represented in figure~\ref{fig:k12}. Note that the spectra obtained from the generated velocity fields show a commendable agreement with those of the DNS data, indicating that the proposed DLM can produce turbulent inflow conditions with a temporal evolution of the velocity fields that is consistent with that of the DNS. \par

\subsection{Interpolation and extrapolation capability of the DLM}

This section investigates the performance of the proposed DLM to generate turbulent inflow conditions at Reynolds numbers that are not used in the training process. The velocity fields at $Re_\theta$ = 763.8 and 1155.1 are used as examples of the velocity fields that fall between the Reynolds numbers used in the training process, i.e. the interpolation ability of the model is investigated using the flow fields at these Reynolds numbers.\par

\begin{figure}
  \centerline{\includegraphics[scale=0.12]{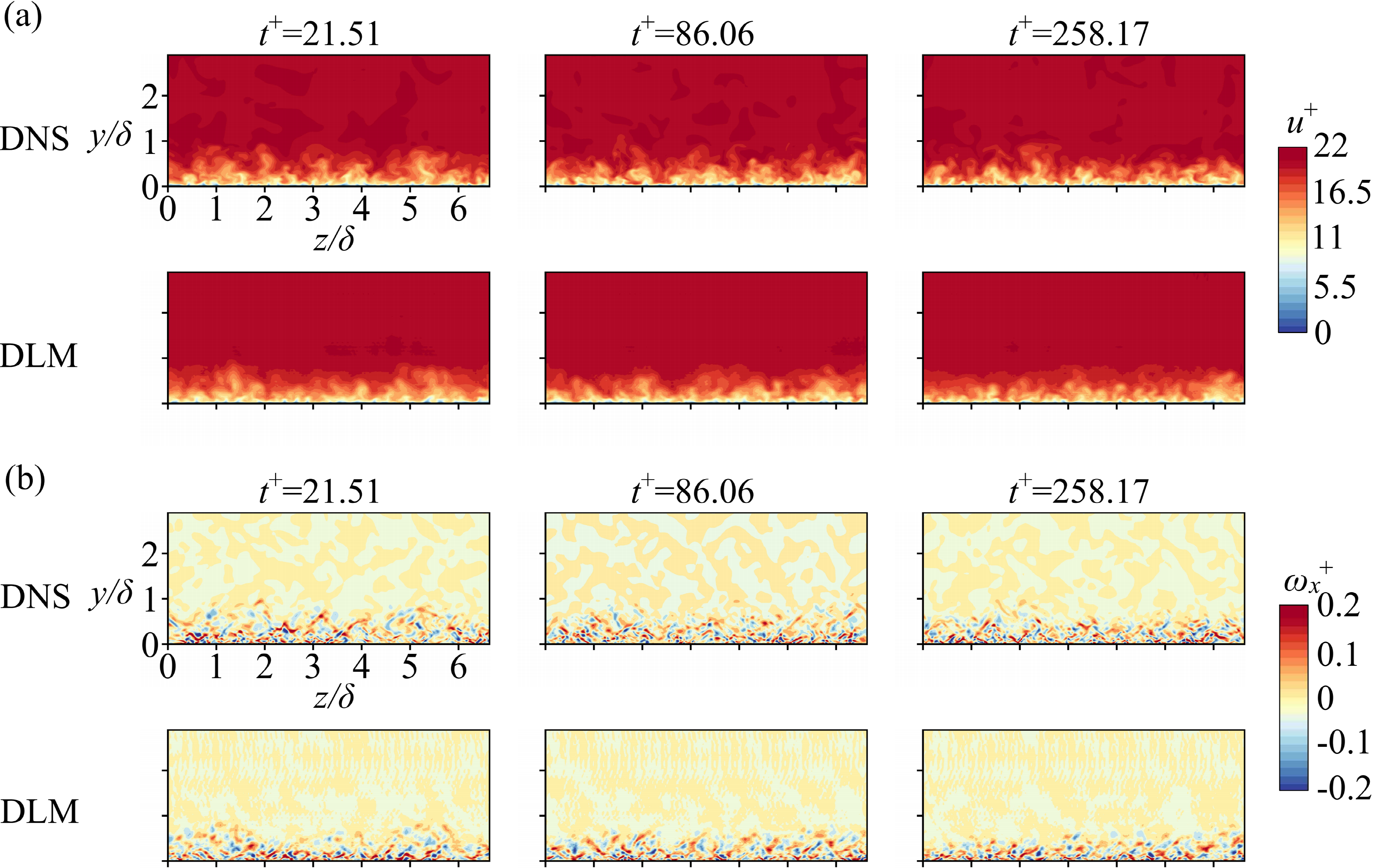}}% Images in 100% size
  \caption{Instantaneous (a) streamwise velocity and (b) vorticity fields at $Re_\theta$ = 763.8, for three different instants. Reference (DNS) and predicted (DLM) data are shown.}
\label{fig:k13}
\end{figure}

\begin{figure}
  \centerline{\includegraphics[scale=0.15]{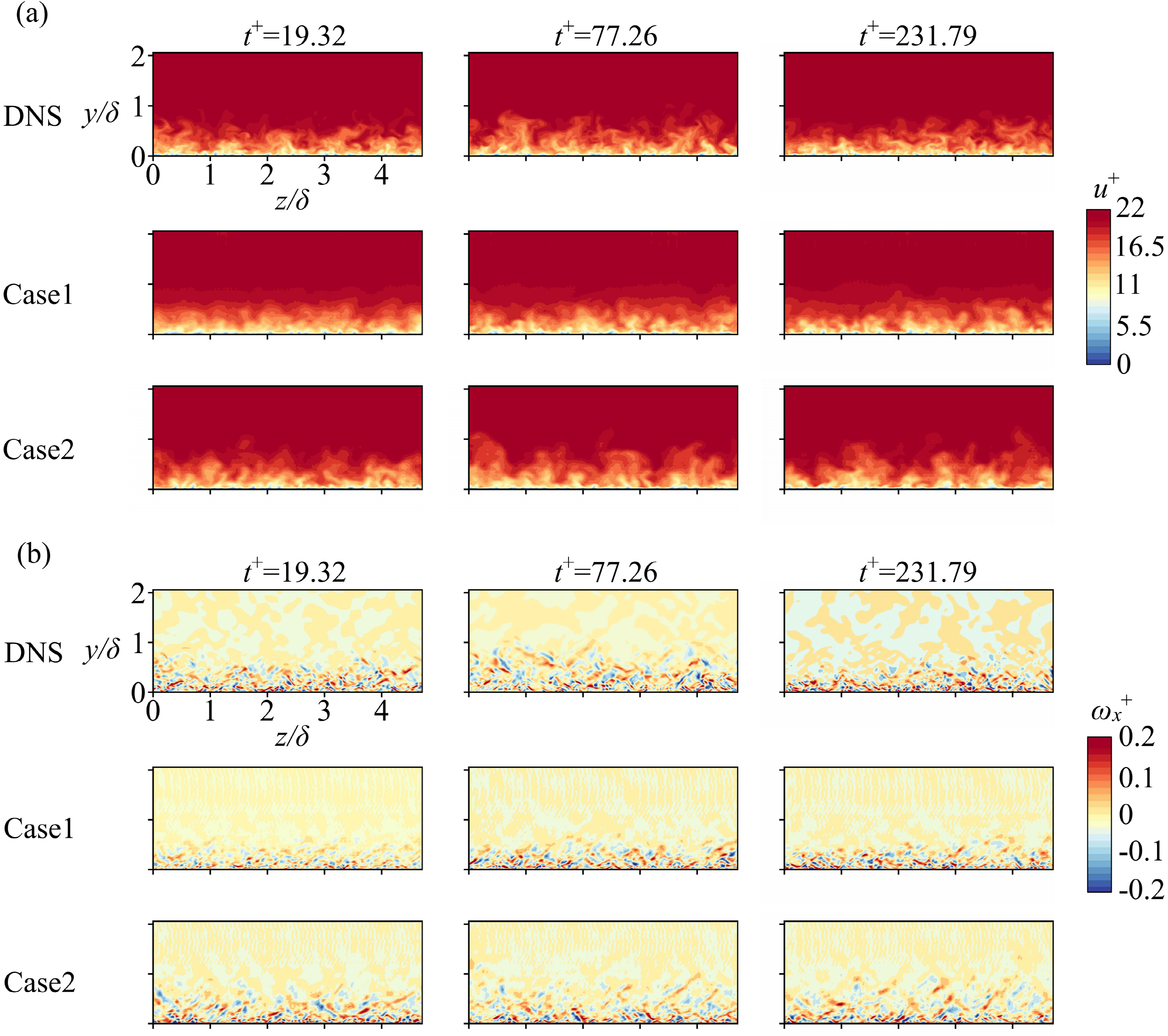}}% Images in 100% size
  \caption{Instantaneous (a) streamwise velocity and (b) vorticity fields at $Re_\theta$ = 1155.1, for three different instants. Cases 1 and 2 represent the prediction using the transformer that is trained for the flow at $Re_\theta$ = 905.7 and 1362.0, respectively.}
\label{fig:k14}
\end{figure}

Figure~\ref{fig:k13} shows the instantaneous streamwise velocity and vorticity fields for the flow at $Re_\theta$ = 763.8. It is worth noting that the transformer trained for the flow at the nearest $Re_\theta$, i.e. $Re_\theta$ = 661.5 is used to predict the temporal evolution of the velocity fields. The figure shows that the main features of the flow fields can be obtained with relatively good precision; however, the details of the predicted velocity fluctuations are not clearly shown. Similar results can be observed in figure~\ref{fig:k14} for the predicted velocity fields at $Re_\theta$ = 1155.1. Here, the current transformers trained for the flow at $Re_\theta$ = 905.7 and 1362.0 are used to predict the temporal evolution of the velocity fields. \par

\begin{figure}
  \centerline{\includegraphics[scale=0.12]{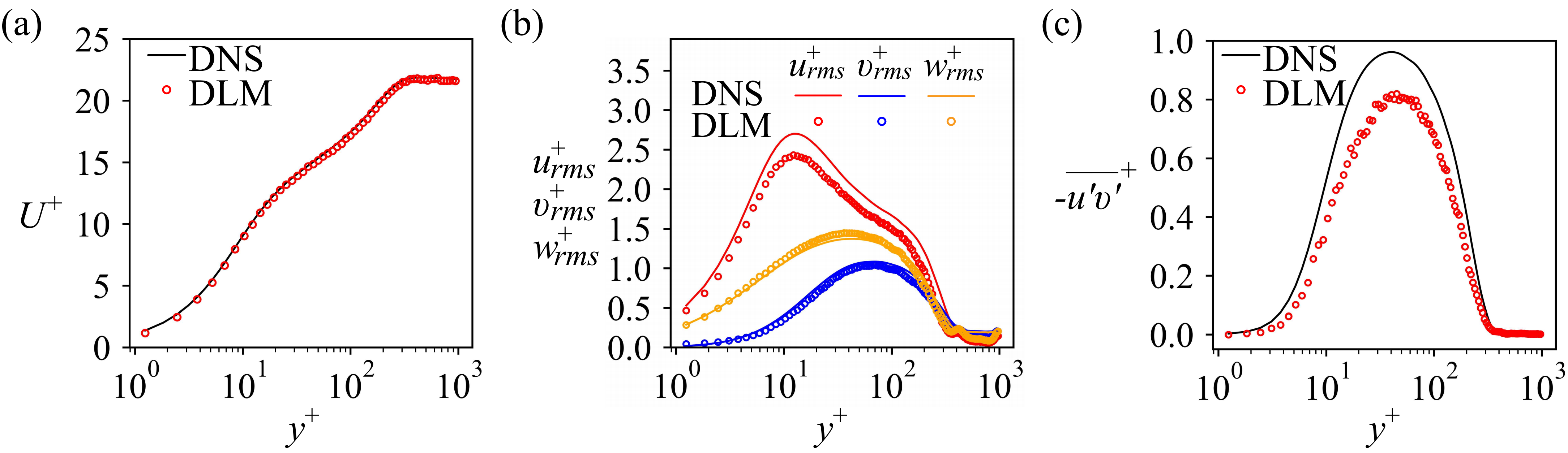}}% Images in 100% size
  \caption{Turbulence statistics of the flow at $Re_\theta$ = 763.8. (a) Mean streamwise velocity profile. (b) Root-mean-square profiles of the velocity components. (c) Reynolds shear stress profile.}
\label{fig:k15}
\end{figure}

\begin{figure}
  \centerline{\includegraphics[scale=0.12]{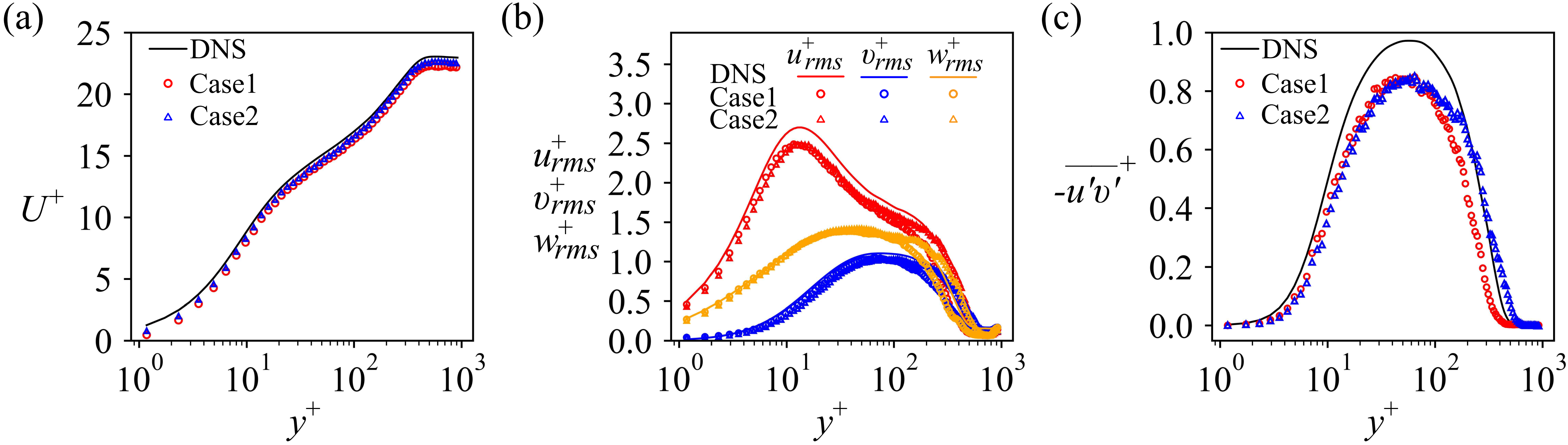}}% Images in 100% size
  \caption{Turbulence statistics of the flow at $Re_\theta$ = 1155.1. Cases 1 and 2 represent the prediction using the transformer model trained for the flow at $Re_\theta$ = 905.7 and 1362.0, respectively. (a) Mean streamwise velocity profile. (b) Root-mean-square profiles of the velocity components. (c) Reynolds shear stress profile.}
\label{fig:k16}
\end{figure}

The turbulence statistics of the flow at $Re_\theta$ = 763.8 and 1155.1 are shown in figures~\ref{fig:k15}  and \ref{fig:k16}, respectively. Although the mean streamwise velocity and the root-mean-square profiles of the spanwise and wall-normal velocity components show an ability of the DLM to predict reasonably well, the root-mean-square profile of the streamwise velocity component and the Reynolds shear stress show an under-prediction due to the lack of detailed information on the velocity fluctuations. 

\begin{figure}
  \centerline{\includegraphics[scale=0.12]{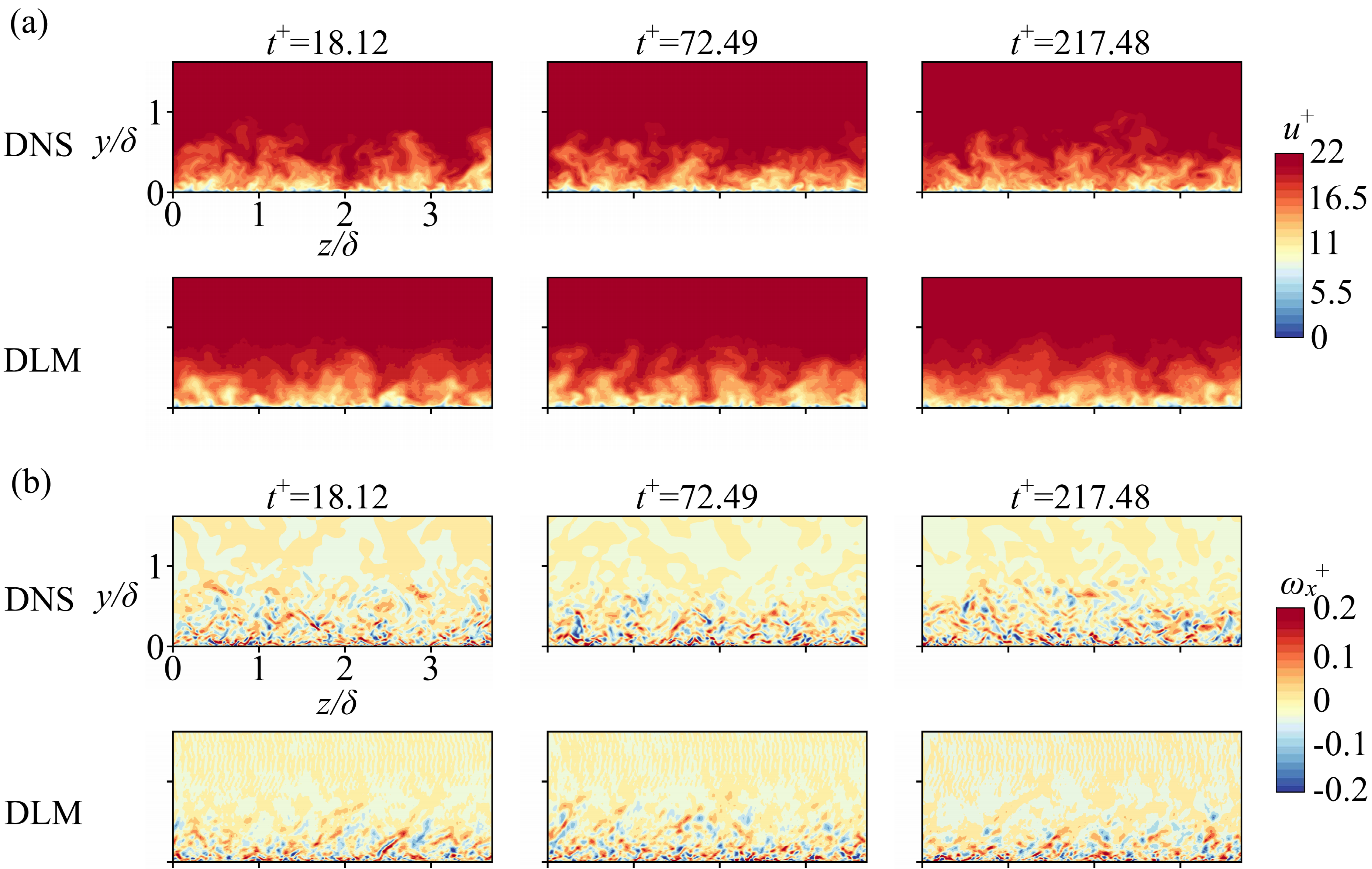}}% Images in 100% size
  \caption{Instantaneous (a) streamwise velocity and (b) vorticity fields at $Re_\theta$ = 1502.0 for three different instants. Reference (DNS) and predicted (DLM) data are shown.}
\label{fig:k17}
\end{figure}

\begin{figure}
  \centerline{\includegraphics[scale=0.12]{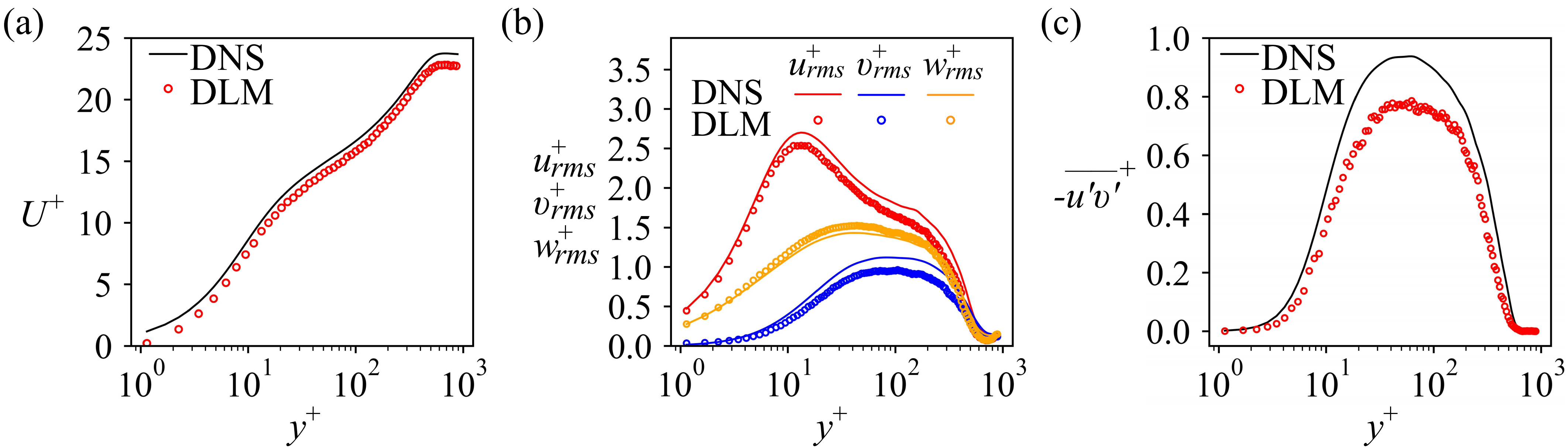}}% Images in 100% size
  \caption{Turbulence statistics of the flow at $Re_\theta$ = 1502.0. (a) Mean streamwise velocity profile. (b) Root-mean-square profiles of the velocity components. (c) Reynolds shear stress profile.}
\label{fig:k18}
\end{figure}

The extrapolation ability of the DLM is evaluated using the flow fields at $Re_\theta$ = 1502.0, which is higher than the maximum $Re_\theta$ used to train the transformer and MS-ESRGAN, i.e. $Re_\theta$ = 1362.0. The transformer trained for the flow at $Re_\theta$ = 1362.0 is used to predict the dynamics of the velocity fields. Figure~\ref{fig:k17} shows that the generated instantaneous streamwise velocity and vorticity fields generally have similar accuracy to the interpolated flow fields. Meanwhile, the turbulence statistics show a deviation from the DNS statistics, as shown in figure~\ref{fig:k18}. This can be attributed to the lack of details of the velocity fluctuations and the extrapolation process that relies on the flow information at one Reynolds number compared with the interpolation process where the flow falls within the range of the Reynolds numbers that the MS-ESRGAN is trained for.\par

\begin{figure}
  \centerline{\includegraphics[scale=0.15]{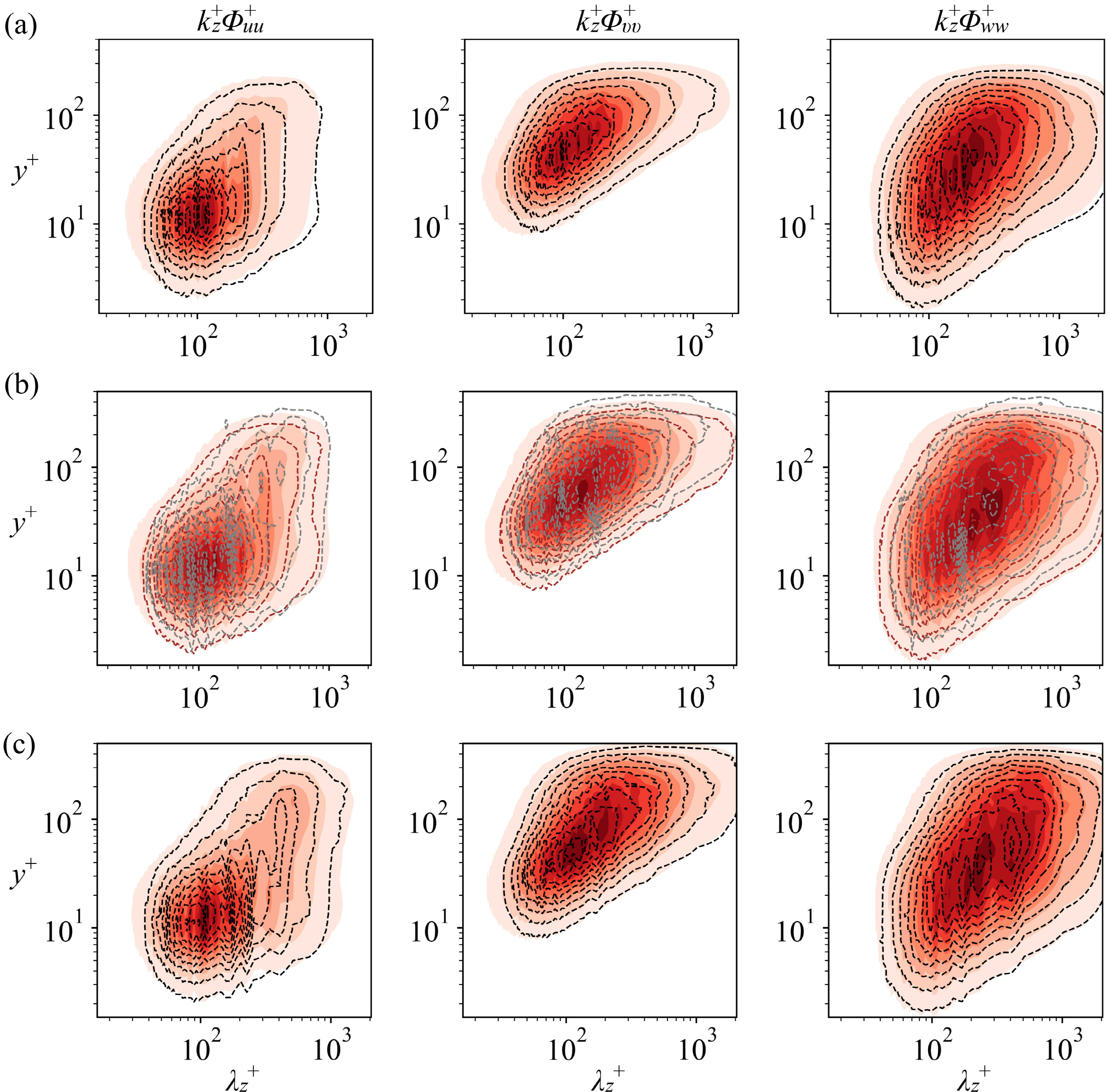}}% Images in 100% size
  \caption{Premultiplied spanwise wavenumber energy spectra of the velocity components as a function of the wall distance and wavelength. The shaded contours represent the results from the DNS data; the dashed-black contours represent the results from the predicted velocity data at $Re_\theta$ = 763.8 and 1502.0; the dashed-brown and grey contours represent the results from the velocity data at $Re_\theta$ = 1155.1 predicted using the transformer model trained for the flow at $Re_\theta$ = 905.7 and 1362.0, respectively. The contour levels are in the range of 10$\%$-90$\%$ of the maximum $k_z^+ \Phi_{\alpha \alpha}^+$ with an increment of 10$\%$.  (a) $Re_\theta$ = 763.8. (b) $Re_\theta$ =1155.1. (c) $Re_\theta$ =1502.0.}
\label{fig:k19}
\end{figure}

Finally, the accuracy of the spectral content of the interpolated and extrapolated velocity components is examined in figure~\ref{fig:k19} by employing the premultiplied spanwise wavenumber spectra. These results indicate that the spectra are produced with relatively good accuracy for the low-moderate wavenumbers.\par

\subsection{Error analysis, transfer learning and computational cost}

The performance of the proposed DLM is further statistically investigated using the $L_2$ norm error of the predicted data for all the Reynolds numbers used in this study,

\begin{equation} \label{eqn:eq11}
\varepsilon = \frac{1}{J} \sum_{j=1}^J  \frac{\left\| {\alpha}_j^{\mathrm {DNS}} - {\alpha}_j^{\mathrm {DLM}}\right\|_2 }{\left\| {\alpha}_j^{\mathrm {DNS}}\right\|_2} 
\end{equation}

\noindent where $\alpha_j^{\mathrm {DNS}}$ and $\alpha_j^{\mathrm {DLM}}$  represent the ground truth (DNS) and the predicted velocity components using the DLM, respectively, and $J$ represents the number of the test snapshots.

\begin{figure}
  \centerline{\includegraphics[scale=0.12]{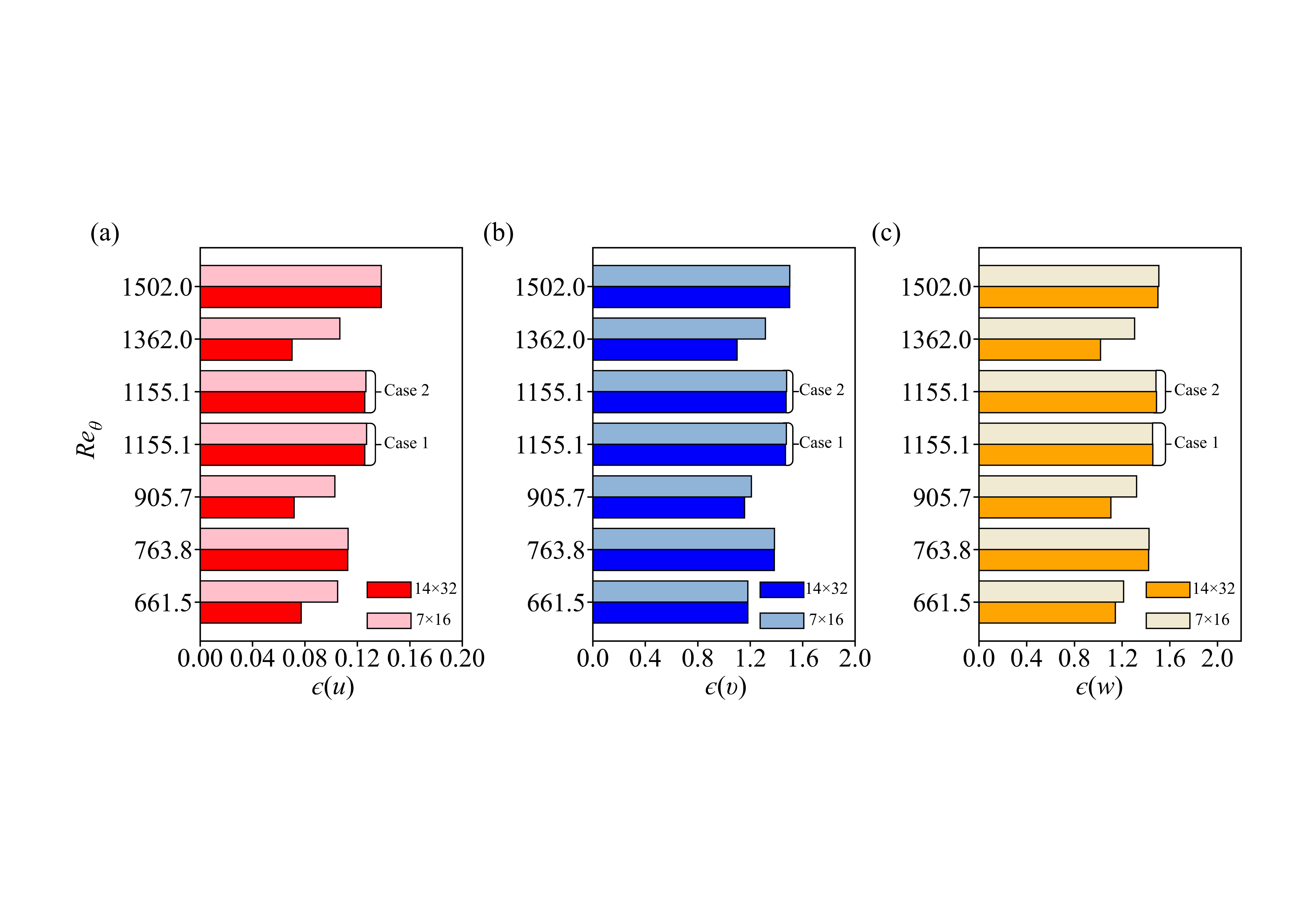}} % Images in 100% size
  \caption{$L_2$ norm error of the predicted velocity fields. (a) Streamwise velocity. (b) Wall-normal velocity. (c) Spanwise velocity. Cases 1 and 2 represent the results from the velocity data at $Re_\theta$ = 1155.1 predicted using the transformer model trained for the flow at $Re_\theta$ = 905.7 and 1362.0, respectively.}
\label{fig:k20}
\end{figure}

Figure~\ref{fig:k20} shows that as the Reynolds number increases, no significant differences can be seen in the error values of the predicted velocity fields. However, as expected, the error shows higher values for the interpolated and extrapolated velocity fields compared with the error of the predicted velocity fields at the Reynolds numbers that the DLM is trained for. Additionally, in contrast with the aforementioned statistical results, the error values are relatively high for the wall-normal and spanwise velocity fields. This indicates that the DLM has learned to model the structure of the flow with generally accurate turbulence statistics and spatio-temporal correlations, rather than reproducing the time sequence of the flow data. This observation is consistent with the results obtained by \cite{Fukamietal2019b, Kim&Lee2020, Yousifetal2022a}. Furthermore, using very coarse input data ($7\times16$) in the training of the DLM shows a slight reduction in the model performance, indicating the capability of the DLM to generate the turbulent inflow data even if it is trained with very coarse input data.

It is worth mentioning that the transfer learning (TL) technique is used in this study \citep{Guastonietal2021,Yousifetal2022a, Yousifetal2021}. The weights of the transformer are sequentially transferred for every training ($y-z$) plane in the flow. First, the transformer is trained for the flow at the lowest Reynolds number, i.e. $Re_\theta$ = 661.5. After that, the weights of the model are transferred for the training using the next $Re_\theta$ data and so on. The results from using TL in this study show that with the use of only 25\% of the training data for the transformer model, the computational cost (represented by the training time) can be reduced by 52\% without affecting the prediction accuracy. These results are consistent with the results obtained by \cite{Guastonietal2021,Yousifetal2022a, Yousifetal2021}.\par

The total number of trainable parameters of the DLM is 356.5 million (305.5 million for the transformer and 51 million for the MS-ESRGAN). The training of the transformer model for all the three Reynolds numbers used in this study using a single NVIDIA TITAN RTX GPU with the aid of TL requires approximately 23 hours. Meanwhile, the training of the MS-ESRGAN requires approximately 32 hours. Thus, the total training time of the DLM model is 55 hours, indicating that the computational cost of the model is relatively lower than the cost of the DNS that is required to generate the velocity fields. Furthermore, this computational cost is required only once, i.e. for the training of the model. Since the prediction process is computationally inexpensive and does not need any data for the prediction (except the initial instantaneous fields), the DLM can also be considered efficient in terms of storing the inflow data. \par

\subsection{Simulation of spatially developing TBL using the turbulent inflow data}

In order to examine the feasibility of applying the DLM-based turbulent inflow conditions, the generated data are utilised to perform an inflow–outflow large eddy simulation (LES) of flat plate TBL spanning $Re_\theta$ = 1362-1820. The open-source computational fluid dynamics (CFD) finite-volume code OpenFOAM-5.0x is used to perform the simulation. The dimensions of the computational domain are 20$\delta_0$, 1.8$\delta_0$, and 4$\delta_0$ in the streamwise, wall-normal and spanwise directions, respectively, where $\delta_0$ represents the boundary layer thickness at the inlet section of the domain. The corresponding grid size =  $320 \times 90 \times 150$. The grid points have a uniform distribution in the streamwise and spanwise directions while local grid refinement is applied near the wall using the stretching grid technique in the wall-normal direction. The spatial spacing at the midpoint of the domain is $\Delta x^+ \approx15.4$,  $\Delta y^+_{wall} \approx0.2$  and  $\Delta z^+ \approx6.5$, where $ y^+ _{wall}$ represents the spatial spacing in the wall-normal direction near the wall. A no-slip boundary condition is applied to the wall, while periodic boundary conditions are applied to the spanwise direction. A slip boundary condition is assigned to the top of the domain, whereas an advection boundary condition is applied to the outlet of the domain. The pressure implicit split operator (PISO) algorithm is employed to solve the coupled pressure momentum system. The dynamic Smagorinsky model \citep{Germanoetal1991} is applied for the subgrid-scale modelling. All the discretisation schemes used in the simulation have second-order accuracy. The generated inflow data are linearly interpolated in time to have a simulation time step $\Delta t = 0.0017 \delta_0/U_\infty$ yielding a maximum Courant number of 0.8. The statistics from the simulation are accumulated over a period of 620 $\delta_0/U_\infty$ after an initial run with a period of 60 $\delta_0/U_\infty$ or 3  flow through.

\begin{figure}
  \centerline{\includegraphics[scale=0.1]{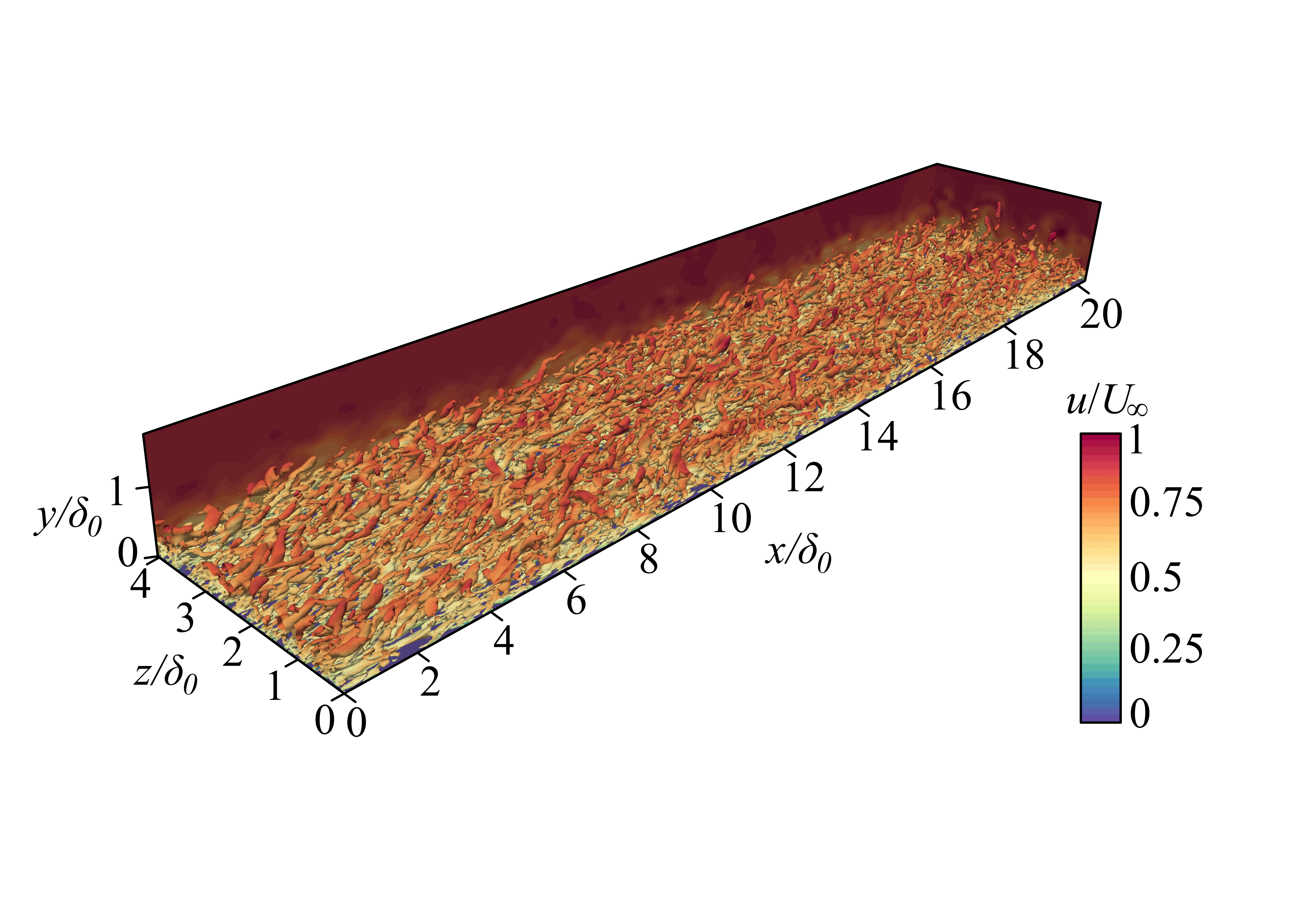}}% Images in 100% size
  \caption{Iso-surfaces of instantaneous vortical structures ($Q-criterion =  0.54U_\infty^2/\delta_0^2$) from the inflow–outflow simulation coloured by the streamwise velocity.}
\label{fig:k21}
\end{figure}

The formation of the instantaneous vortical structures of the flow is visualised by utilising the $Q-criterion$ vortex identification method \citep{Huntetal1988} in figure~\ref{fig:k21}. Smooth development of the coherent structures represented by the hairpin-vortex-like structures \citep{Adrian2007} can be observed from the figure with no noticeable formation of artificial turbulence at the inlet section of the domain. This indicates that the inflow data obtained from the DLM could represent most of the flow physics at the inlet section, resulting in a negligible developing distance upstream of the domain.

\begin{figure}
  \centerline{\includegraphics[scale=0.12]{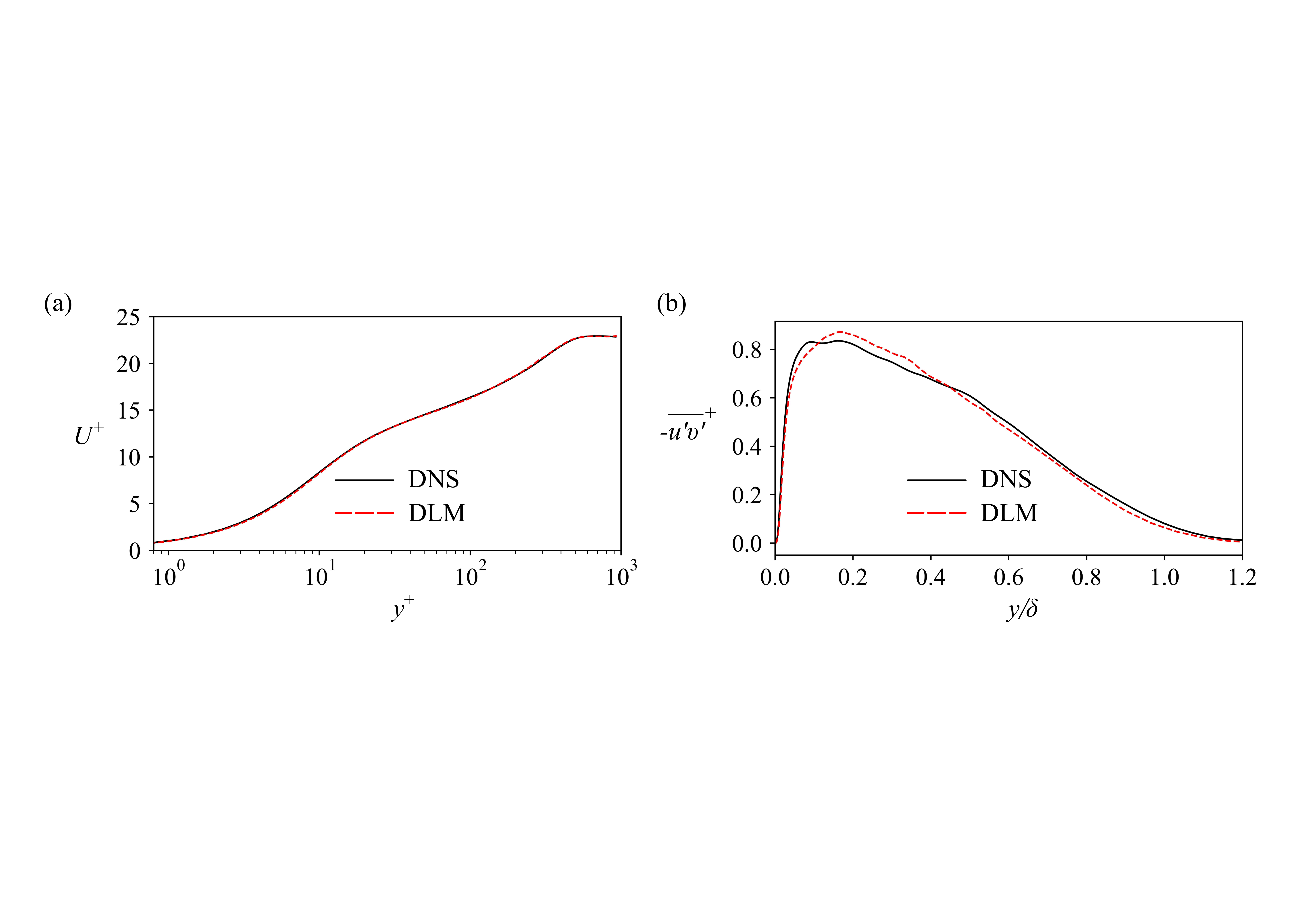}}% Images in 100% size
  \caption{Turbulence statistics from the inflow–outflow simulation at  $Re_\theta$ = 1400 compared with the DNS results. (a) Mean streamwise velocity profile. (b) Reynolds shear stress profile.}

\label{fig:k22}
\end{figure}

A comparison of the mean streamwise velocity and Reynolds shear stress profiles at $Re_\theta$ = 1400 with the DNS results are provided in figure~\ref{fig:k22}. The mean streamwise velocity profile is in excellent agreement with the DNS results. Furthermore, the Reynolds shear stress profile is consistent with the DNS results in most of the boundary layer regions. 

\begin{figure}
  \centerline{\includegraphics[scale=0.12]{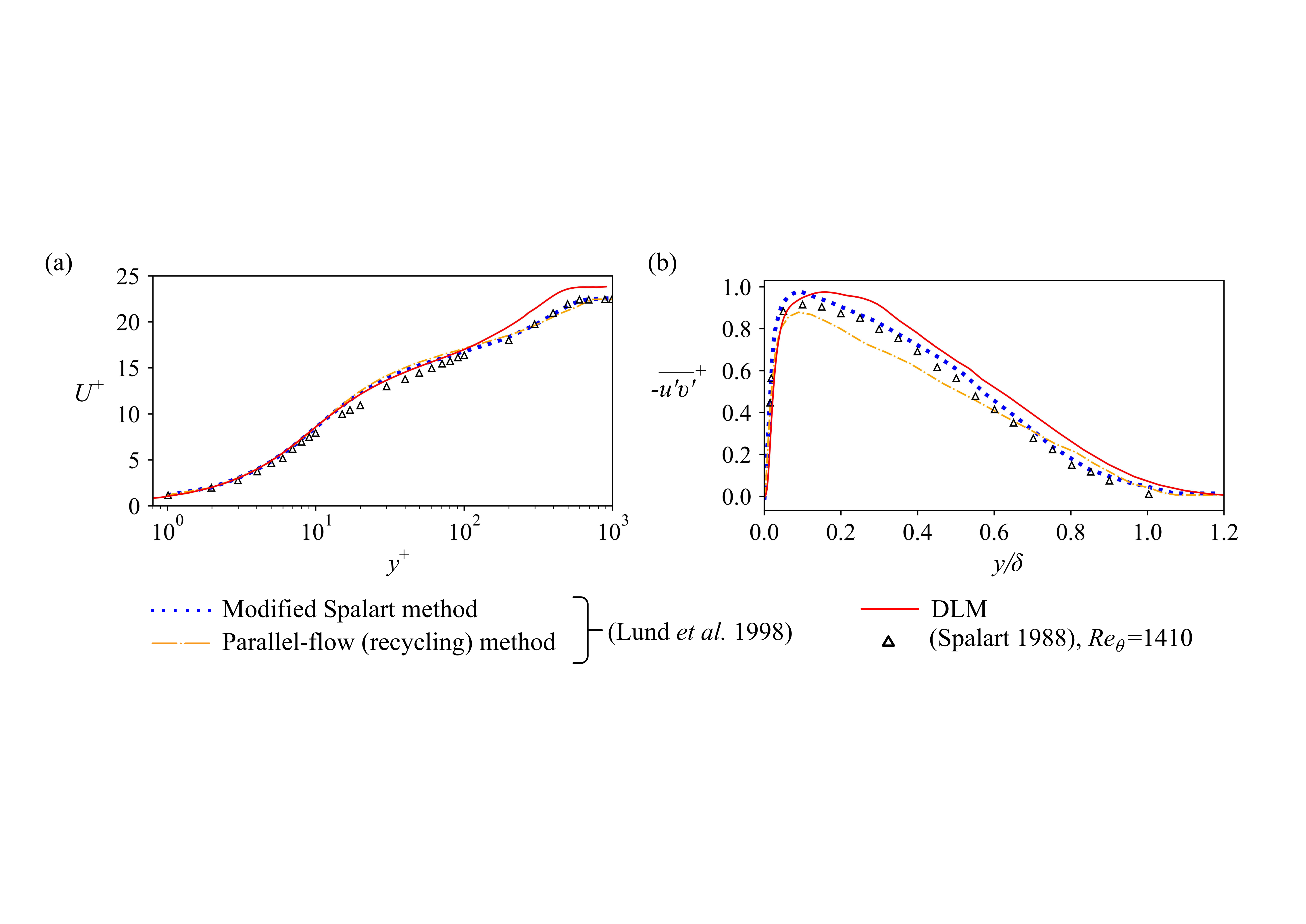}}% Images in 100% size
  \caption{Turbulence statistics from the inflow–outflow simulation at  $Re_\theta$ = 1530 compared with the results of \cite{Lundetal1998} and \cite{Spalart1988}. (a) Mean streamwise velocity profile. (b) Reynolds shear stress profile.}
\label{fig:k23}
\end{figure}

To further evaluate the accuracy of the inflow conditions, statistics obtained from the simulation are compared with the inflow–outflow LES results of \cite{Lundetal1998} and DNS results of  \cite{Spalart1988}. Figure~\ref{fig:k23} shows the profiles of the mean streamwise velocity and Reynolds shear stress profiles at $Re_\theta$ = 1530. An agreement can be observed with \cite{Lundetal1998} results of the modified Spalart (recycling-rescaling) method and the results of \cite{Spalart1988} ($Re_\theta$ = 1410 ) in the inner region of the boundary layer, however, a deviation can be observed in the outer region. This might be attributed to the fact that the original DNS data that are used to train the DLM contain free-stream turbulence \citep{Lee&Zaki2018}. 

\begin{figure}
  \centerline{\includegraphics[scale=0.12]{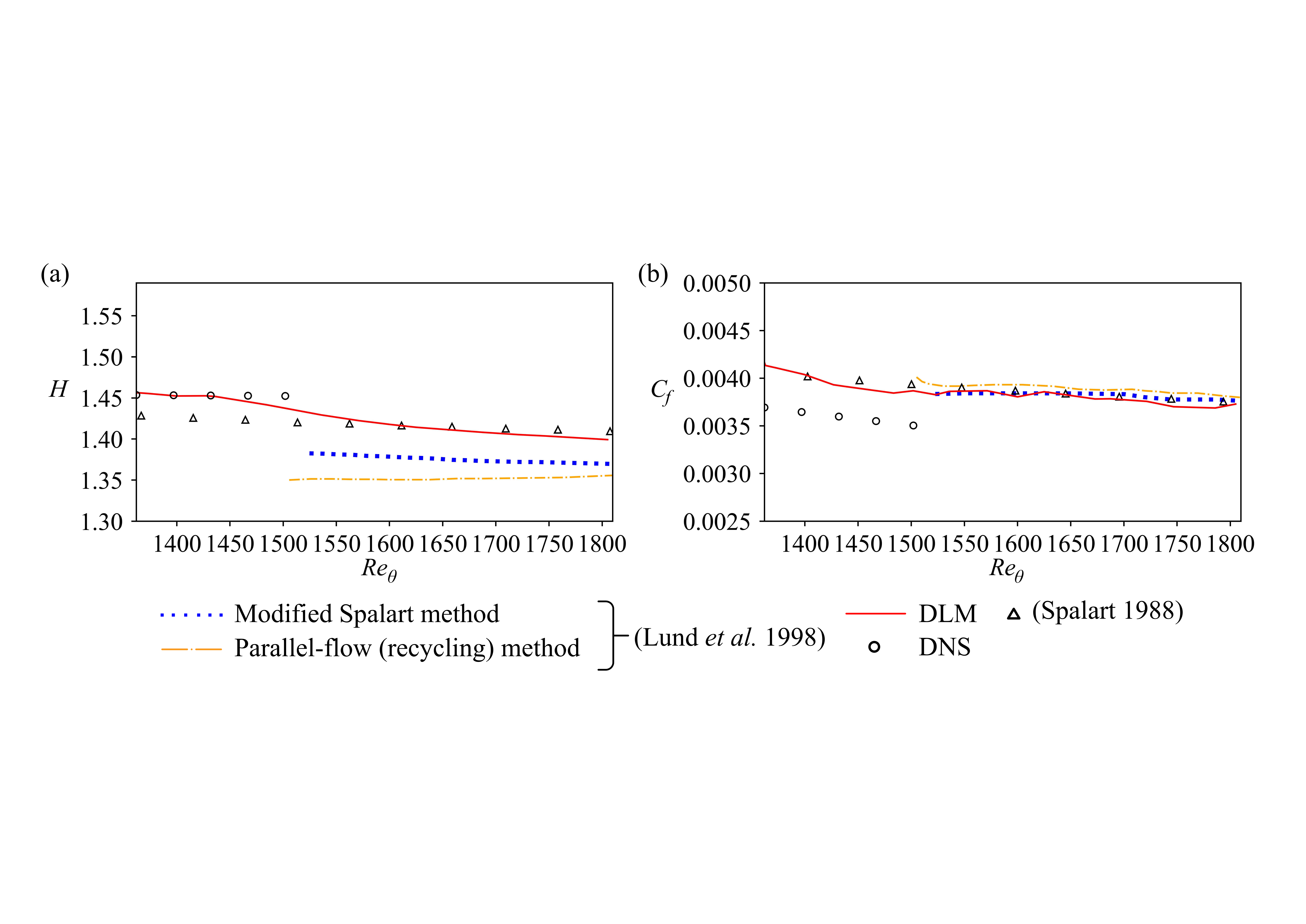}}% Images in 100% size
  \caption{Evolution of the shape factor and skin-friction coefficient in the inflow–outflow simulation compared with the results of the DNS, \cite{Lundetal1998} and \cite{Spalart1988}. (a) Shape factor. (b) Skin-friction coefficient.}
\label{fig:k24}
\end{figure}

The evolution of the shape factor $H$ is shown in figure~\ref{fig:k24}(a). Here the result from the simulation is generally consistent with the DNS and \cite{Spalart1988} results, and within $5\%$ of the modified Spalart method from  \cite{Lundetal1998}. The result of the skin-friction coefficient in figure~\ref{fig:k24}(b) shows an agreement with the results from \cite{Lundetal1998} and \cite{Spalart1988} with an over-prediction of approximately $8\%$ compared with the DNS results. The change in the shape factor slope and the over-prediction of the skin-friction coefficient compared with the DNS result can be attributed to the numerical setup of the inflow-outflow simulation. Note that in the work of \cite{Lundetal1998}, the inflow data were generated from precursor simulations that have the same $y-z$ plane size as the inlet section of the inflow-outflow simulations, and no spatial or time interpolation was applied to the inflow data.

The above results suggest that the turbulent inflow data that are generated by the proposed DLM can be practically used as inflow conditions for simulations that do not necessarily have the same spatial and time resolutions as the generated data, which is the case in the simulation described in this section.

\section{Conclusions} \label{sec:5}
 
This study proposed a deep-learning-based method to generate turbulent inflow conditions for spatially-developing TBL simulations. A combination of a transformer and MS-ESRGAN was used to build the inflow generator. The transformer was trained to model the temporal evolution of the velocity fields represented by various ($y-z$) planes of spatially limited data. Meanwhile, MS-ESRGAN was trained to perform super-resolution reconstruction of the predicted velocity fields.\par

The generated instantaneous velocity fields showed an excellent agreement with the DNS results for the velocity fields at Reynolds numbers that the DLM was trained for. The model also successfully reproduced the turbulence statistics with commendable accuracy. Furthermore, the model reproduced the spectra of the velocity components with accurate precision, indicating accurate spatial and temporal correlations of the generated velocity components, which further supports the ability of the model to maintain the realistic behaviour of the velocity fields.\par

The performance of the proposed model was further examined using velocity fields at Reynolds numbers that were not used in the training process. The instantaneous and statistical results showed a reasonable accuracy for the interpolated and extrapolated velocity fields. The spectra of the velocity components revealed a relatively good agreement with the results from the actual velocity data, with a deviation that can be observed at high wavenumbers. These results suggest that the model can generate the turbulent inflow conditions for the flow at Reynolds numbers that are not necessarily used in the training of the model.\par

The results obtained from the error analysis showed that the increase in the Reynolds number has no significant effect on the error values of the predicted velocity fields, indicating that the model is robust to the increase of the Reynolds number. The use of TL in the training of the transformer revealed a noticeable reduction in the computational cost of the DLM without affecting the precision of the prediction. \par

The inflow-outflow simulation results showed the feasibility of applying the generated turbulent inflow conditions to turbulent flow simulations as a negligible developing distance upstream of the domain is required for the TBL to reach the target statistics.

This study showed for the first time that a transformer-based model could be effectively used for modelling the dynamics of turbulent flows with the ability to perform parallel computing during the training process, which is not possible in LSTM-based models. It also paved the way for utilising synthetic-inflow generators for large-scale turbulence simulations using deep learning, with significant promise in terms of computational savings. \par

\section*{Acknowledgements}
This work was supported by 'Human Resources Program in Energy Technology' of the Korea Institute of Energy Technology Evaluation and Planning (KETEP), granted financial resource from the Ministry of Trade, Industry \& Energy, Republic of Korea (no. 20214000000140). In addition, this work was supported by the National Research Foundation of Korea (NRF) grant funded by the Korea government (MSIP) (no. 2019R1I1A3A01058576). This work was also supported by the National Supercomputing Center with supercomputing resources including technical support (KSC-2021-CRE-0244). R.V. acknowledges the financial support from the ERC Grant No.
“2021-CoG-101043998, DEEPCONTROL”.

\section*{Declaration of interests}
The authors report no conflict of interest.

% Susie put cite commands here, don't bother with sites, etc just yet.

\bibliographystyle{jfm}

\bibliography{JFM2022}

\end{document}